Magneto-Transport and Spin-Reorientation in Pt/$Co_{78}Ho_{22}$ Heterostructures Near the Sublattice Compensation Temperature

Rajeev Nepal[1], Jose Flores[2], Aurain Seaton[1], Michael Newburger[2], John Derek Demaree[3] and Ramesh C. Budhani[1*]

[1]Department of Physics, Morgan State University, Baltimore, Maryland 21251, USA.

[2]Materials and Manufacturing Directorate, Air Force Research Laboratory, Dayton, Ohio 45433, USA.

[3]DEVCOM Army Research Laboratory, Aberdeen Proving Ground, Aberdeen, MD, 21005, USA.

**Abstract**

Metallic amorphous ferrimagnets of 3*d* transition metals (TM) and rare earths (RE) with *4f* electrons exhibit rich magneto-transport behavior due to the interplay between the 3*d* and 4*f* magnetic sublattices and their interaction with mobile charges. Tuning the TM and RE concentrations in the alloy can effectively modulate the compensation temperature, where the moments of the two sublattices point in opposite direction leading to a net zero magnetization. Despite extensive magnetotransport studies in Gd- and Tb-based 3*d*–4*f* systems, Ho-based alloys remain comparatively underexplored, even though Ho possesses the largest orbital angular momentum (OAM) among the lanthanides. This unquenched OAM can strongly impact magnetic anisotropy and magnetotransport in ferrimagnetic heterostructures. Here, we have investigated the anomalous Hall resistivity ($\rho_{xy}^{AHE}$), dc magnetization, and spin Hall magnetoresistance (SHMR) of $Co_{78}Ho_{22}$/Al film and a Pt/$Co_{78}Ho_{22}$/Al heterostructure deposited using multitarget magnetron sputtering. The $\rho_{xy}^{AHE}$ of both systems shows a distinct sign reversal and prominent wing-shaped hysteresis loops in the vicinity of the compensation temperature ($T_{comp}$), which is accompanied by the minimum saturation magnetization near $T_{comp}$. Furthermore, the SHMR in Pt/$Co_{78}Ho_{22}$/Al film is enhanced due to the Pt layer. These HM interface-induced prominent features of magneto-transport are addressed in the light of the existing theories of spin–flop transitions, spin orbit torque, and microscopic phase separation, which may lead to the formation of 3*d* and 4*f* magnetic clusters in the film.

## I. INTRODUCTION

Ferrimagnetic spintronics is rapidly emerging as an exciting field of research [1-2], leveraging materials that combine antiferromagnetic-like order with ferromagnetic tunability [3]. In rare-earth (RE) transition-metal (TM) ferrimagnets (FEMs), the opposing sublattice magnetizations can cancel at a specific composition or temperature, resulting in nearly zero net magnetization. This so-called compensated ferrimagnetic state [3] offers several advantageous features like those of pure antiferromagnets, including negligible stray magnetic fields and insensitivity to external magnetic field perturbations, which are important for energy-efficient spintronics [4–8]. However, understanding and controlling the magnetic order in FEMs remains challenging, particularly in the vicinity of the compensation point. A major issue is the internal competition between the two magnetic sublattices, which manifests in several complex ways. Notably, spin reorientation transitions (SRTs) or spin-flop (SF) phenomena may occur, marked by a breakdown of collinearity between the sublattice magnetizations [9]. In ferrimagnets containing heavy rare-earth elements like Dy and Ho with large 4*f* orbital moments, these transitions are often first-order in nature, driving the system from a collinear ferrimagnetic state to a non-collinear or "sperimagnetic" spin configuration on approaching the compensation temperature [10,11]. Such rich phase transitions are accompanied by the formation of non-trivial magnetic textures such as stripes or bubble domains, often associated with SRTs [12]. Theoretical phase diagrams indeed suggest that the relative anisotropy of the RE vs. TM sublattices, as well as interface-induced anisotropy in thin films, are key factors governing these spin reorientation transitions [6, 11, 13].

Some RE–TM ferrimagnetic thin films with perpendicular magnetic anisotropy exhibit a wing-shaped anomalous Hall resistivity $\rho_{xy}^{AHE}$ (H) loops [14-17], and magneto-optical response [18-24] in the vicinity of the compensation temperature ($T_{comp}$). Such materials include thin films of HoCo [14], TbFe [15], GdFe [16, 18], DyCo [19], GdFeCo [9, 20, 22], and GdCo [23]. Under normal circumstances, applying a perpendicular magnetic field should cause a ferrimagnet's net magnetization to increase and align with the field direction. However, in these RE–TM films, a counterintuitive response is observed: the Hall and Kerr signals drop above a certain field magnitude and can even nearly disappear [15,16]. This peculiar high-field behavior underscores the complex interplay between the two magnetic sublattices in RE–TM ferrimagnets around the compensation point [9, 21]. While such triple-step or "wing-shaped" hysteresis loops near the $T_{comp}$ have been extensively investigated, the physical origin of this phenomenon is a topic of debate. For instance, one school of thought attributes it to a spin-flop transition between antiferromagnetically coupled 3d and 4f magnetic sublattices once the Zeeman energy exceeds the inter-sublattice exchange, leading to a field-induced reorientation into a canted state [9,11,16]. An alternative explanation invokes subtle compositional inhomogeneities or nanoscale RE–TM phase separation that creates locally compensated regions with staggered switching fields [14, 23]. Despite these pioneering works, the identification of a precise mechanism for the formation of the wing-shaped loops still warrants further theoretical and experimental efforts. Triple hysteresis loops in RE–TM ferrimagnets have been widely studied via electrical transport (anomalous Hall effect) [14-17] and magnetization-based methods such as DC magnetometry [18], X-ray Magnetic Circular Dichroism (XMCD) [19,20], and the magneto-optical Kerr effect [21-23]. Most prior work employed only one technique at a time, while only a few combined multiple approaches [15]. In this work, we integrate simultaneous magnetization and transport measurements on the same samples, directly correlating the material's magnetic state with its Hall response through the triple-loop regime.

The heterostructures of RE-TM films with heavy metals (HM) such as Ta, Pt, Ir, W, etc., have also garnered much interest in addressing the spin Hall effect-related phenomena at the interface [25-33]. Such HM/FEM heterostructures may also exhibit magnetic proximity effects, whereby a small magnetic moment is induced in the nominally nonmagnetic HM layer at the interface [31]. This proximity-induced magnetization (PIM) is highly sensitive to the electronic structure of the heavy metal; specifically, in the metals with high 4d or 5d-electron density near the Fermi level, such as Pd, Pt, Ir, or W [6, 34-36]. In contrast, heavy metals with less favorable electronic configurations, such as filled or poorly hybridizing 5d bands, exhibit negligible PIM [37,38]. For instance, gold (Au), with a filled 5d band, and tantalum (Ta) with a partially filled 5d band do not readily become spin-polarized when interfaced with a magnetically ordered material due to their lower Stoner factor and unfavorable 5d band structure [37-40]. Since the PIM moment in a RE-TM system will align parallel to the magnetic moment of the TM sublattice due to its delocalized 3d wavefunction, it may shift the compensation temperature. The PIM layer may also lead to spin–orbit torque (SOT) on the ferrimagnetic layer in the presence of thermal gradients, light fields, and RF and dc currents. The efficiency of such a process is enhanced as the net magnetization diminishes near the compensation point [25, 26].

Given a strong spin-orbit coupling (SOC) in Pt, heterostructures of Pt with TM-RE films are expected to show rich SOT and PIM-related effects. Amongst the various RE–TM ferrimagnetic alloys, the Gd-TM systems have been studied extensively, revealing rich interfacial magnetic and electronic phenomena ascribed to the Pt layer [26-31]. However, this problem is expected to become much more interesting when a Pt overlayer is deposited on a RE-TM alloy film made of a rare-earth element with stronger SOC. Recent measurements show that the lanthanide, holmium (Ho), exhibits the strongest SOC as well as the largest orbital angular momentum, while others like Gd or Lu display much weaker contributions [41]. Earlier works have mainly focused on the high SOC elements like Tb and Dy [42, 43], but the FEM alloys of the element Ho have remained relatively understudied [44]. The HoCo alloy films are of particular interest as they show a strong magneto-crystalline anisotropy [45]. This extreme anisotropy, combined with the antiparallel coupling of Ho and Co sublattices, makes Ho–Co an ideal prototype to investigate temperature-driven spin reorientation transitions, spin pumping, and magneto-transport anomalies around the compensation point. Notably, there have been very few efforts to address the effects of magnetic compensation on electronic transport and magnetic ordering in Ho-Co films. Previous studies on Ho–Co alloys have reported perpendicular magnetic anisotropy and anomalous Hall effect sign reversals near the compensation temperature [14, 46], along with temperature-driven spin reorientation and related magnetotransport anomalies [44]. However, these results are based on isolated measurements, and a comprehensive understanding of the magnetic and transport behavior in HoCo thin films has not yet been established. Moreover, the transport phenomena such as spin Hall magnetoresistance (SMR), anisotropic magnetoresistance (AMR), and AHE remain unaddressed in strong SOC metal / HoCo heterostructures. In particular, the influence of Pt interfacing on the magnetic compensation behavior, spin-flop regime, and correlated magneto-transport response in HoCo systems has not been systematically investigated. Clearly, a dedicated study of such phenomena in a RE-TM alloy where the RE element carries the largest orbital angular momentum is warranted. Such a study would potentially establish whether spin currents selectively probe one magnetic sublattice when the net moment vanishes, the nature of SMR as the system passes through compensation, and the role of the large orbital angular momentum of Ho atoms in these processes.

Here, we address these questions by combining AHE, SMR, AMR, and DC magnetization measurements on sputter-deposited $Co_{78}Ho_{22}$ thin films with and without a Pt underlayer. We show that Pt interfacing not only modifies the compensation temperature and enhances the net magnetization but also broadens the triple-loop (spin-flop) regime and markedly alters the magnetotransport response in the vicinity of compensation. By varying the temperature across the $T_{comp}$, a distinct emergence of a wing-shaped, triple-step hysteresis in the AHE loops is seen, and correlated with a field-induced, sublattice-specific spin-flop transition that shows as an inflection point in the magnetization curves. Furthermore, the SMR increases while the orbital magnetoresistance (OMR) is suppressed, underscoring the selective sensitivity of spin-current transport to the sublattice. These results highlight the sublattice magnetization switching and interfacial spin scattering that govern magneto-transport in a scantly studies, high sublattice angular momentum ferrimagnet in HoCo, positioning this material as a novel system for understanding compensated ferrimagnetism in spintronic architectures.

## II. EXPERIMENTAL DETAILS

Thin films of composition $Co_{78}Ho_{22}$ and Pt/$Co_{78}Ho_{22}$ were deposited at ambient temperature onto amorphous quartz substrates using DC magnetron co-sputtering of high-purity elemental cobalt and holmium targets in a confocal geometry under an argon pressure of 5 mTorr. The depositions were carried out in a load-locked, all-metal-sealed vacuum chamber with a base pressure of 8 x $10^{-9}$ Torr. The Co, Ho and Pt targets were powered at 160 W, 32 W and 25 W, respectively, yielding deposition rates of 0.67 Å/s for Co, 0.47 Å/s for Ho and 0.4 Å/s for Pt, as determined from prior calibration measurements. The elemental composition of the films was nominally controlled by these calibrated deposition rates and the known material densities and confirmed by Rutherford Backscattering (RBS) and earlier by X-ray photoelectron spectroscopy [44]. For the RBS analysis, a relatively high incident $He^+$ energy of 4.365 MeV was employed to clearly resolve the Ho and Pt signals. Data fitting and interpretation were carried out using SIMNRA, which incorporates corrections for detector resolution, energy straggling, interfacial roughness, and thickness variations across the ~1 mm RBS beam spot. RBS data and analysis are provided in Fig. S1 of the Supplemental Information.

X-ray diffractometry (XRD) measurements were performed with a Rigaku SmartLab X-Ray diffractometer using a 4-bounce monochromator. X-ray reflectivity (XRR) measurements were analyzed using SmartLab Studio simulations to obtain the layer thicknesses of the heterostructure. Atomic force microscopy (AFM) and Magnetic force microscopy (MFM) measurements were conducted using a Bruker Dimension Icon 5 in tapping mode and lift mode, respectively. MFM scans were performed at a height of 25 nm using magnetic Multi75M cantilevers.

To study the effects of the heavy-metal overlayer on the magnetic and magneto-transport properties, a second set of $Ho_{22}Co_{78}$ (henceforth labeled as HoCo) films was grown on a 9 nm-thick Pt film, which was deposited first *in situ* at 25 W power with a rate of 0.4 Å/s. In both sets of samples, a 3 nm-thick aluminum (Al) capping layer was deposited over the HoCo layer to prevent surface oxidation. All films were patterned into Hall bar geometries using shadow masks, enabling measurements of AHE, AMR, and SMR. Magnetization and magnetotransport measurements were performed in a physical property measurement system equipped with a 9 T superconducting solenoid, a vibrating sample magnetometer attachment, and a sample rotator, allowing both in-plane and out-of-plane rotation of the sample with respect to the direction of the applied magnetic field. These measurements were carried out over a wide temperature range (10 – 300 K) to probe the changes in magnetic and transport properties across the compensation temperature.

## III. RESULTS AND DISCUSSION

This section begins with a description of the structural and magnetic characterization of films. It is followed by a detailed analysis of magnetization and Hall resistivity measurements to determine the compensation temperature and to examine the enhancement in saturation magnetization in Pt/Ho–Co heterostructures, indicative of interfacial proximity effects. The emergence of the triple-loop anomalous Hall resistivity and angular magnetoresistance is discussed next, which reveals a sublattice-dependent switching behavior and persistent SMR response near compensation, reflecting selective coupling of spin currents to the Co sublattice.

### III.a Structural and Magnetic Characterization:

The crystallographic structure, interface quality, surface topography, and magnetic domain patterns of the CoHo/Pt bilayers have been probed with a combination of X-ray diffraction, XRR, AFM, and MFM techniques. The X-ray reflectivity of the film measured with CuKα radiation, shown in Fig. 1(a), reveals clear oscillations from the platinum layer and smaller oscillations from the HoCo layer, from which the thicknesses of individual layers were calculated. The result of X-ray diffraction measurements on the CoHo/Pt bilayer is presented in Fig. 1(b). The scattered intensity profile is characterized by a broad hump centered at 2Θ ~ $22^{0,}$ indicating the presence of a non-crystalline material, along with weak reflections corresponding to the scattering of X-rays from the (111) platinum plane. No distinct peaks from HoCo are observed, except for the broad intensity profile at 2Θ ~ $22^{0}$, consistent with the amorphous or nanocrystalline structure typically associated with thin films of rare-earth transition-metal alloys deposited at ambient temperature.
The AFM image shown in Fig. 1(c) reveals a uniform and finely textured surface morphology of surface roughness ≈ 0.86 nm, which suggests a sharp interface between HoCo and Pt layers. Further, the film does not show large topographic features or rough spikes, which may damage the smoothness of the interface locally. Fig. 1(d) presents the corresponding MFM image, which shows a dense labyrinthine domain pattern composed of alternating out-of-plane magnetic domains, as evidenced by the bipolar contrast. These stripe-like domains form spontaneously at remanence and indicate strong perpendicular magnetic anisotropy in the HoCo film. Notably, there is no apparent correlation between domain positions and topographic features, suggesting that domain formation is primarily driven by intrinsic magnetic properties and not by surface roughness or microstructural inhomogeneities.

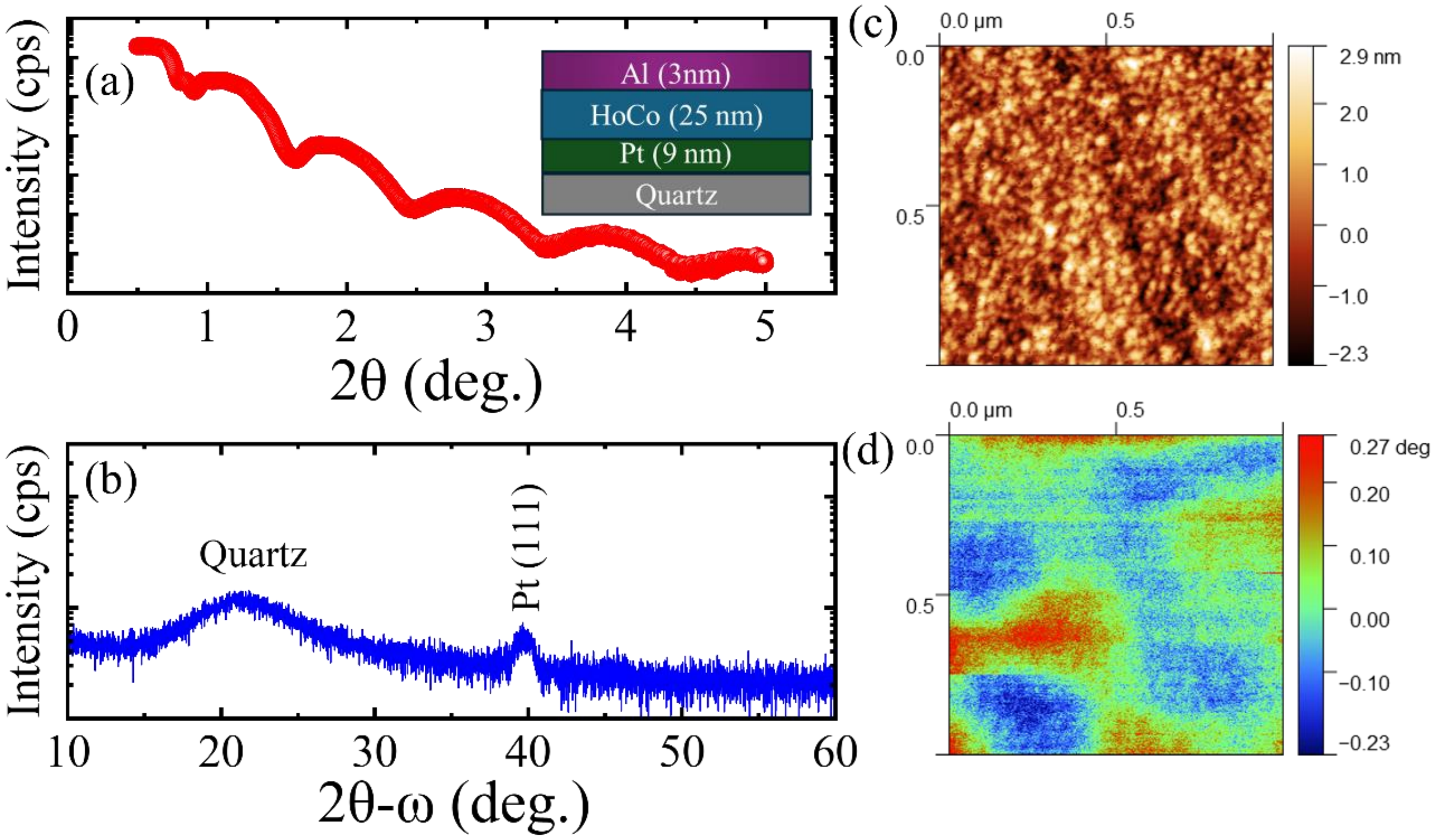

*Figure 1: Topographical and magnetic textures of a 25 nm HoCo film deposited on a 9 nm Pt underlayer on an amorphous quartz substrate. (a) XRR profile with pronounced interference fringes, indicating a smooth, continuous film of sharp interfaces. (b) XRD pattern showing a broad amorphous background from the quartz substrate and weak Pt (111) reflection. (c) AFM image showing a smooth surface with fine nanoscale texture with an average roughness of ≈ 0.86 nm. (d) MFM image revealing a dense labyrinthine domain pattern, characteristic of out-of-plane magnetic anisotropy in ferrimagnetic films.*

III.b: Magnetic Compensation and Interface Effects in HoCo/Pt Heterostructures

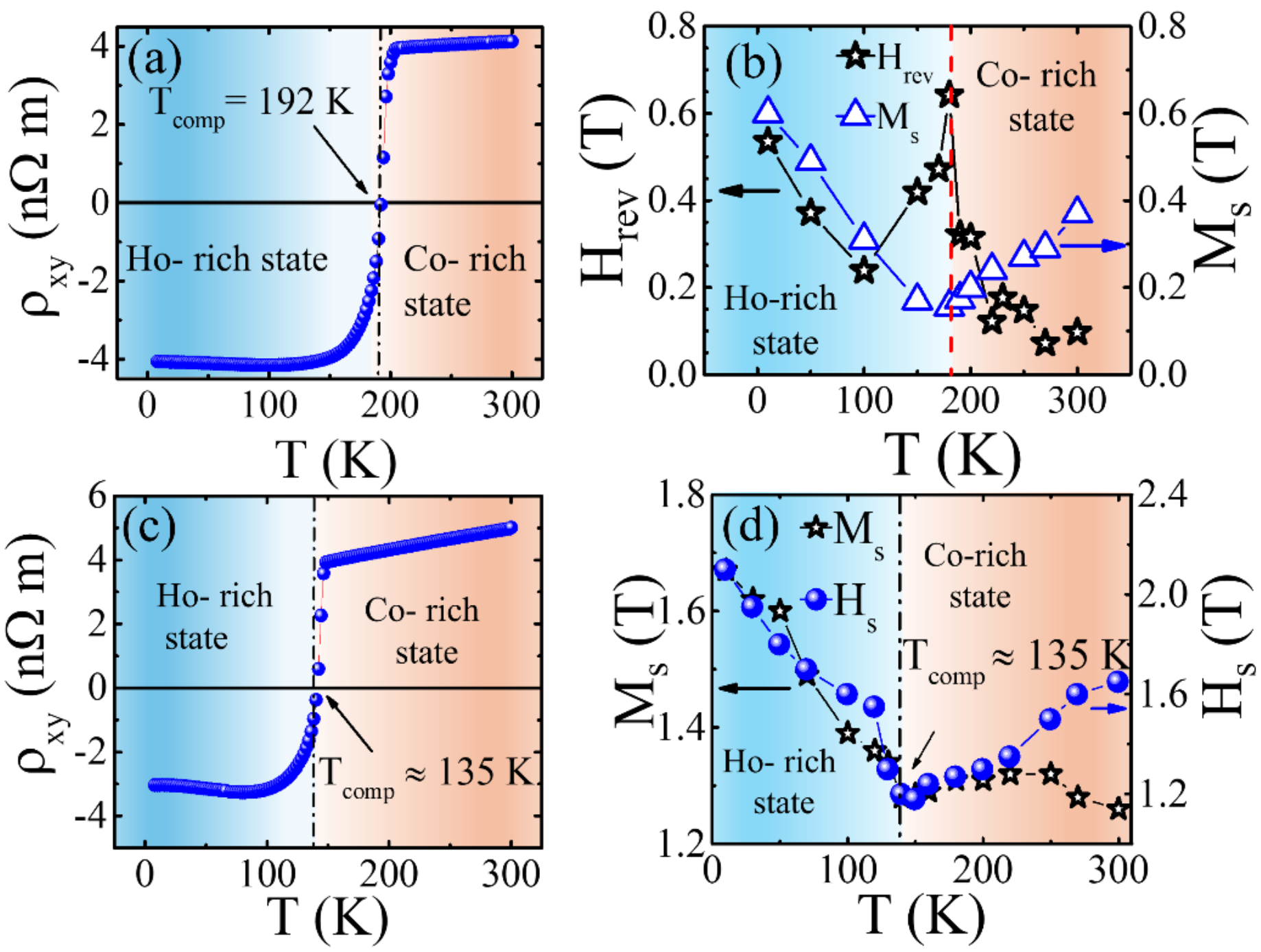

*Figure 2: Magnetization and Hall resistivity measurements on HoCo film and HoCo/Pt bilayer, with the magnetic field applied perpendicular to the film plane (out-of-plane). (a) Hall resistivity $\rho_{xy}$ as a function of temperature of the bare 25 nm thick HoCo film measured at 2 tesla, showing a polarity reversal at the magnetic compensation temperature ($T_{comp}$) ~ 192 K. All $\rho_{xy}$ data have been antisymmetrized to isolate the odd Hall component. (b) Temperature dependence of the saturation magnetization ($M_s$) and the magnetization reversal field ($H_{rev}$) for the HoCo film, with $M_s$ nearly vanishing at $T_{comp}$. (c) $\rho_{xy}$ of the HoCo/Pt bilayer at 2 tesla, exhibiting a sign reversal at ~ 135 K. (d) Temperature dependence of $M_s$ and the saturation field ($H_s$) for the HoCo/Pt bilayer, demonstrating a markedly enhanced $M_s$ across the entire temperature range compared to that of the bare HoCo film. Blue and orange shaded regions denote the Ho-dominated and Co-dominated magnetic states, respectively.*

Figures 2(a) to 2(d) present a comparative analysis of the Hall resistivity and dc magnetization of HoCo and Pt/HoCo thin films. The saturation magnetization ($M_s$), reversible field ($H_{rev}$), and saturation field ($H_s$) in Fig. 2 were extracted from the M(H) loops shown in Figs. S2 and S3 of the Supplementary Information. In Fig. 2(a), the $\rho_{xy}$ exhibits a clear sign reversal at T ≈ 192 K, which is a signature of the compensation temperature. In 3d ferromagnets like Co, the ($\rho_{xy}^{AHE}$) is positive, whereas elemental Ho exhibits a negative $\rho_{xy}^{AHE}$ in its magnetically ordered state [44]. This difference in the sign of AHE implies that at $T < T_{comp}$ the magnetization of the Ho sublattice ($\mathbf{M}_{Ho}$) is parallel to the applied field direction, whereas the $\mathbf{M}_{Co}$ is antiparallel. In contrast, a positive $\rho_{xy}$ above $T_{comp}$ reflects the dominance of the Co-sublattice magnetization ($\mathbf{M}_{Co}$). The saturation magnetization of the film becomes minimum at $T_{comp}$, as seen in Fig. 2(b), indicating that the $\mathbf{M}_{Ho}$ and $\mathbf{M}_{Co}$ are nearly equal and opposite in sign. This behavior is characteristic of ferrimagnetic RE-TM systems where the two antiferromagnetically coupled sublattices have different thermal demagnetization rates, leading to a compensation temperature at which their opposing moments balance each other and the net magnetization essentially vanishes. Noticeably, the reversibility field ($H_{rev}$), defined as the field beyond which the magnetization is fully reversible in M(H) loops, goes through a peak at the compensation temperature. This is indicative of a frustrated regime of magnetization with memory that refused to saturate at small fields. This regime of temperature also supports stabilization of magnetic textures such as skyrmions and bubbles [8].

Interestingly, the Hall resistivity and magnetization characteristics of the HoCo/Pt bilayer, presented in Figs. 2 (c) and 2 (d), respectively, reveal a significantly different behavior. The change in the sign of $\rho_{xy}$ now occurs at a lower temperature of 135 K (Fig. 2(c)), coinciding with the minimum in $M_s$ at the same temperature (Fig. 2(d)). Importantly, there is also a two-fold increase in the value of saturation magnetization when the HoCo film is deposited over the layer of platinum. The downward shift in $T_{comp}$ seen in Fig. 2(c) is presumably due to the interfacial magnetic proximity effect and compositional asymmetry introduced by the Pt underlayer by creating a Ho-deficient, Co-rich interface with reduced Ho magnetization, as similarly reported in other Pt-capped ferrimagnetic alloys such as Pt-FeCoGd-Pt trilayers [31]. In essence, the heavy-metal Pt acquires a proximity-induced magnetic moment that couples to the Co sublattice at the interface, as previously reported by Swindells et al through resonant X-ray scattering measurements [31]. Here, the induced moment in Pt aligns with the transition-metal sublattice moment rather than the rare-earth moment. This interfacial asymmetry effectively increases the Co-like contribution to the magnetization at the interface, meaning a larger $\mathbf{M}_{Ho}$ is required to achieve compensation, hence a lower $T_{comp}$. Furthermore, the proximity-induced magnetic moment in Pt also introduces an anomalous contribution to its Hall resistivity. This proximity-induced AHE has been observed in Pt/ferromagnet and Pt/ferrimagnet bilayers, where it follows the interfacial magnetization in both coercivity and loop shape, increases in magnitude for ultrathin Pt layers due to its interfacial origin, and is strongly suppressed when a nonmagnetic spacer such as Cu interrupts the exchange coupling [47].

Concomitant with a changed $T_{comp}$, the Pt interface also enhances the net magnetization of the bilayer, as shown in Fig. 2(d), where the $M_s$ of the Pt/HoCo bilayer reaches ≈ 1.63 Tesla, compared to ≈0.38 Tesla for the bare HoCo film at 300 K. Notably, even at the $T_{comp}$ where the bare HoCo film exhibits a nearly vanishing $M_{s,}$ the Pt/HoCo structure maintains a substantial net magnetization of ≈1.2 Tesla. While this four-fold enhancement in moment could suggest PIM in the Pt layer, a simple estimate based on bulk magnetometry yields an unrealistically high average moment of ~1.5 μB per Pt atom, which is well above the typical XMCD-reported values of ≈ 0.3 μB per atom in Pt/CoGd and Pt/FeCoGd systems [31]. This discrepancy implies that the observed increase in magnetization is likely to include contributions beyond PIM, such as interfacial exchange effects [48] or modifications to the HoCo magnetization profile. We further note that a second Pt/HoCo film with a compensation temperature near room temperature exhibits a similar enhancement in magnetization (Supplementary Information Fig.S5), indicating that this behavior is robust and reproducible rather than specific to a single composition.

One likely mechanism to account for this additional magnetization is strong orbital hybridization at the Pt/HoCo interface, which may occur between Pt *5d* and Co *3d* orbitals and lead to a narrowing of the Co *3d* band and an enhancement of its local magnetic moment. This mechanism seems operational in Co/Pt multilayers, where the XMCD measurements have revealed that such interfacial hybridization not only enhances the orbital magnetic moment of Co but also drives perpendicular magnetic anisotropy through Co–Pt orbital mixing [49]. Further, our Pt underlayer exhibits a pronounced (111) texture, as confirmed by X-ray diffraction. This orientation is known to enhance Co–Pt orbital hybridization, thereby favoring a stronger PIM moment in the Pt layer [50]. Element-specific measurements such as XMCD at the Pt $L_{2,3}$ edges would be valuable for directly resolving the interfacial magnetic contributions, which could be a scope for future studies.

In summary, interfacing HoCo with a Pt underlayer dramatically reconfigures its magnetic state, yielding a pronounced downward shift in the compensation temperature and a strong enhancement of the net magnetization. These interface-driven modifications underscore the potential of heavy-metal/ferrimagnet heterostructures for spintronic device engineering, where tuning the compensation point and boosting magnetization can be leveraged for robust operations.

III. c. Triple-Loop Hall Effect and Sublattice Reversal in Ho–Co Ferrimagnets

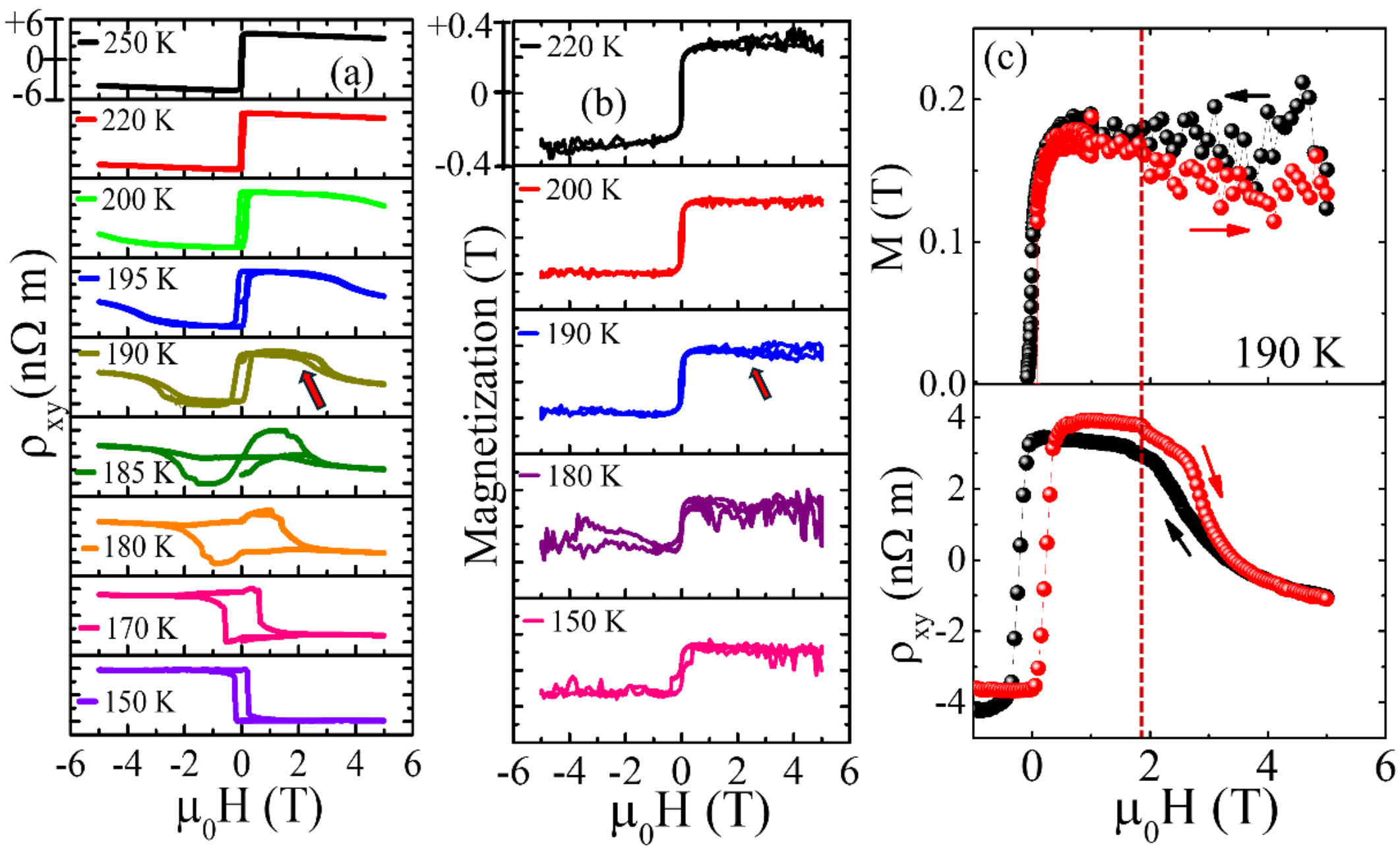

*Figure 3: (a) Hall resistivity ($\rho_{xy}$) of HoCo film over a temperature range of 150 to 250 K. A dramatic change in the shape of the $\rho_{xy}$ loops, along with a sign reversal are seen in the figure. A distinct triple-loop (pointed by the red arrow) in the $\rho_{xy}$ (H) is observed near the compensation temperature of $T_{comp} \approx 190$ K. (b) Corresponding out-of-plane magnetization (M–H) loops of the same sample at T = 150, 180, 190, 200, and 220 K. The red arrow in Fig. 3(b) is suggestive of a metamagnetic spin-flop (SF) transition. (c) Comparison of positive field branches (0 to +5 T) of $\rho_{xy}$ and M(H) loops at T = 190 K, highlighting the difference in magnetization reversal behavior observed in transport and magnetization. The dashed line in Fig. 3(c) shows the SF transition occurring around the spin-flop field ($\mu_0 H_{SF}$) ~ 2 tesla in both $\rho_{xy}$ and M(H) measurements.*

Figure 3(a) shows the AHE loops of the bare HoCo film measured at various temperatures. At T ≤ 185 K, the AHE is negative. However, as the temperature approaches $T_{comp} \approx 190$ K, the AHE vs field plot undergoes a marked evolution in shape, developing pronounced asymmetry and exhibiting a triple hysteresis loop. The Hall resistivity at $\mu_0 H > 3$ tesla is close to zero. While the second loop in AHE disappears at $T \geq 195$ K, the signature of a drop in AHE at higher field remains till T = 200 K for a maximum field of 5 tesla. The observation of triple loops in AHE as the magnetic field is scanned between the positive and negative quadrants has been a topic of great debate in the RE-TM alloy literature [14-24, 51, 52]. Two main schools of thought have emerged regarding the origin of these triple loops. In one scenario, microscopic compositional variations such as slight rare-earth enrichment in certain regions lead to spatial gradients in $T_{comp}$ across the film [14, 17, 18, 23]. The material effectively partitions into exchange-coupled sub-layers with different switching fields, so the measured hysteresis is a superposition of partial loops from these regions [14, 18, 51]. An alternative explanation invokes a field-induced spin-flop transition between the antiferromagnetically coupled 3d and 4f sublattices when the applied field is near $T_{comp}$ [9, 16]. In this picture, once the Zeeman energy from the external field overcomes the inter-sublattice exchange coupling, the two sublattices abruptly reorient into a canted, noncollinear configuration, such as in a first-order metamagnetic transition that produces a sudden magnetization reversal at a critical field [9]. Theoretically, if the anisotropy of the rare-earth sublattice is sufficiently large, a spin-flop phase boundary can persist even above $T_{comp}$, giving rise to wing-shaped loops in a *homogeneous* ferrimagnet without invoking chemical inhomogeneity [11]. Apart from these primary mechanisms, several secondary factors have been proposed. For instance, nanoscale phase separation into RE-rich and TM-rich regions can create an internal exchange-bias-like effect, which was reported to yield multi-step loops and extremely large bias fields in Dy–Co films [19, 20, 24]. The proposed mechanism here is a nanoscale compositional phase separation into Dy-rich and Co-rich regions within a single film, creating internal exchange-coupled interfaces that behave like an intrinsic exchange-bias system [20,24]. This nanoscale RE/TM segregation produces distinctly multi-step hysteresis loops and exceptionally large loop shifts (bias fields) in Dy–Co films, essentially acting as an internal pinning field [24]. By contrast, the earlier "phase separation" scenarios [14, 17, 18] refer to more conventional coexisting-phase effects at a larger scale.

Our measurements on bare HoCo thin films provide direct evidence of two-stage sublattice switching near the $T_{comp}$ ($\approx$190 K), supporting the intrinsic spin-flop scenario. As shown in Fig. 3(b), the M(H) exhibits a narrow central hysteresis loop at low fields, followed by a weak metamagnetic transition at approximately 2 tesla, marked by an inflection in the slope. Simultaneously, the anomalous Hall resistivity $\rho_{xy}$ at 190 K displays a triple-loop profile characterized by two distinct reversal steps: a sharp jump near zero field, and a broader transition at higher field ~ 2 tesla. These results can be understood in the framework of the two-sublattice Hall effect mechanism proposed by Okamoto and Miura for ferrimagnetic RE–TM films [16]. In this scenario, the Hall voltage in the field regime of the first loop is an additive sum of the transition-metal and rare-earth sublattice contributions $V_H(I) = |V_H|(RE) + |V_H|(TM)$, since the two moments are antiparallel and carry opposite Hall polarities. Under a sufficiently strong magnetic field, a spin-flop transition may occur that forces both sublattice magnetizations to align with the field (a high-field "phase II" state), in which the Hall resistivity is given by the difference of the two sublattice contributions $V_H(II) = |V_H|(TM) - |V_H|(RE)$. Our HoCo films exhibit analogous behavior. We examine the behavior of this apparent spin reorientation transition further by comparing M(H) and $\rho_{xy}$(H) data at 190 K on an expanded scale in Fig. 3(c). As evident from the shape of the second loop of $\rho_{xy}$(H) at 190 K, the reorientation in the forward and reverse branches of the loop does not have the same field dependence, leading to a sizable hysteresis.

Further, on completion of the reorientation, the $\rho_{xy}$ is distinctly negative. Based on the behavior of the $\rho_{xy}$(H), we surmise that it is the $\mathbf{M}_{Ho}$ that is undergoing a rotation from antiparallel to parallel configuration with respect to $\mathbf{M}_{Co}$ on increasing the field. Further, the rotation angle θ has a different field dependence on the increasing and decreasing field branches of the loop. This will account for the observed hysteresis. While this intuitive picture explains the behavior of the $\rho_{xy}$(H), it must also be consistent with the nature of the M(H) curve. However, the changes in the magnetization are minimal, although it does show a hysteretic behavior in the field regime of the second loop of the $\rho_{xy}$(H). Near the $T_{comp}$ (≈190 K), the asymmetric features observed in the $M(H)$ loops likely arise from field-induced canting and reorientation of the antiferromagnetically coupled Ho and Co sublattices, which modify the net ferrimagnetic moment through unequal sublattice contributions. In contrast, the anomalous Hall response exhibits a comparatively more symmetric multi-step switching behavior because $\rho_{xy}$(H) is predominantly governed by the transition-metal sublattice and remains highly sensitive to sublattice-specific spin-dependent scattering and reorientation processes near the spin-flop-like transition [16]. A likely scenario under which the behaviors of $\rho_{xy}$(H) and M(H) could be reconciled is that the $\mathbf{M}_{Ho}$ is much smaller than the $\mathbf{M}_{Co}$, but it does contribute to the Hall resistivity to account for the negative value of $\rho_{xy}$ on completion of the reorientation.

Importantly, this multi-step switching is observed only in the vicinity of $T_{comp}$, where $M_{net}$ is close to zero. Away from $T_{comp}$, the loops revert to single-step behavior as one sublattice dominates. The sensitivity of this response to both field and temperature confirms that HoCo hosts a finely balanced two-sublattice structure near compensation, in which modest fields can tip the magnetic equilibrium and reveal spin-reversal pathways otherwise hidden in conventional ferrimagnets [54, 55]. Taken together, our M(H) and Hall measurements on HoCo most likely constitute strong experimental evidence for intrinsic, anisotropy-governed spin-flop transitions as the source of triple-loop behavior in this ferrimagnet.

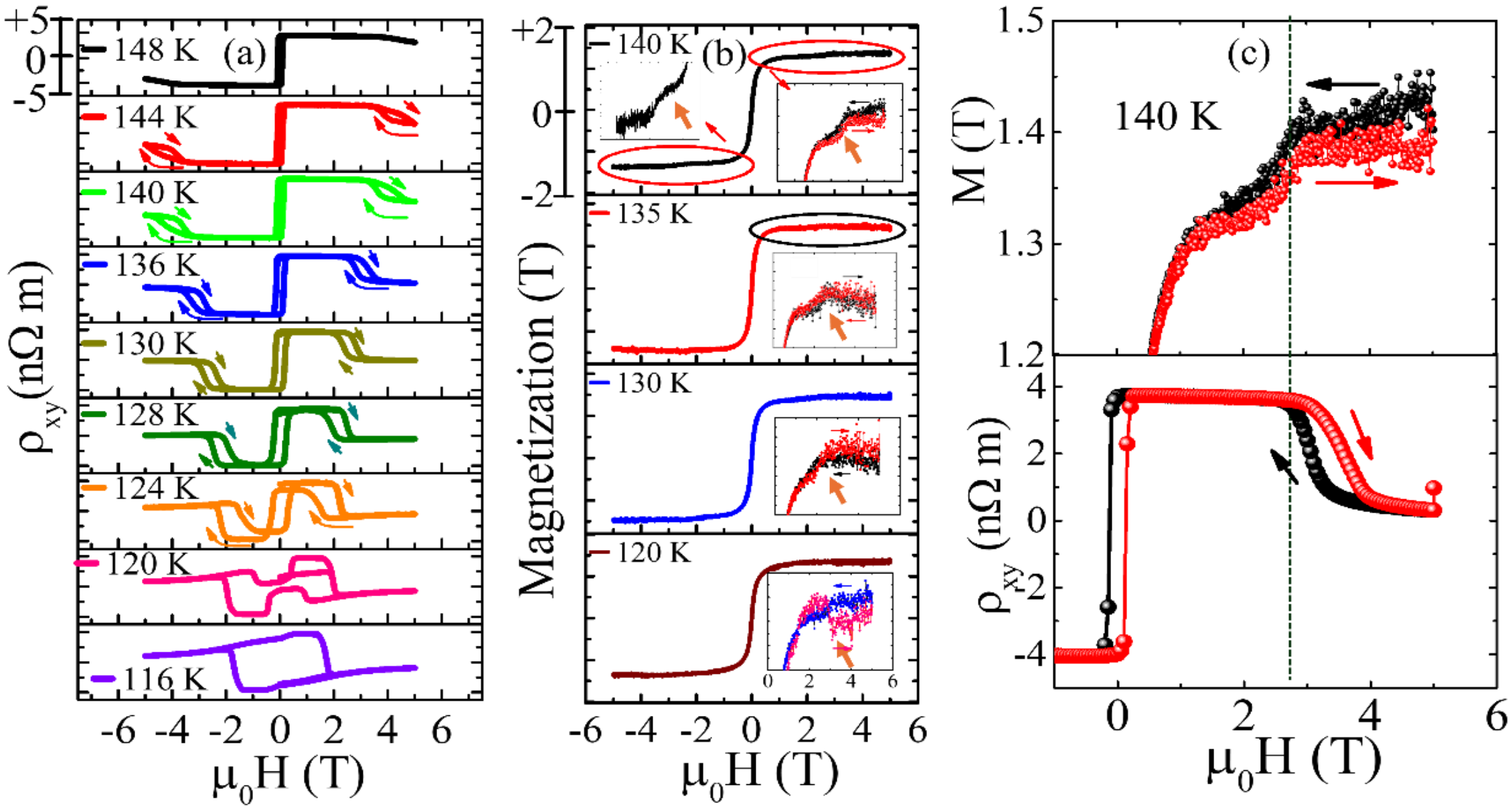

*Figure 4: (a) Anomalous Hall hysteresis loops ($\rho_{xy}$(H)) of Pt/Ho–Co films measured across temperatures from 116 to 148 K. A distinct triple-loop structure is observed around the compensation temperature of ≈ 135 K. (b) Corresponding out-of-plane magnetization (M(H))*

*loops for the same sample at temperatures 120, 130, 135, and 140 K. Insets highlight the intermediate-field region where metamagnetic transition occurs (Pointed by the arrows), consistent with a field-induced spin-flop–like transition between Ho and Co sublattices. (c) Comparison of positive field branches (0 to +5 T) of $\rho_{xy}$ and M(H) loops at T = 140 K, highlighting the difference in sublattice magnetization reversal behavior observed in transport and magnetization. The dashed line in Fig. 4(c) shows the SF transition occurring around the spin-flop field ($\mu_0 H_{SF}$) ~ 2.6 T in both $\rho_{xy}$ and M(H) measurements.*

Figure 4 (a) shows the AHE loops of the HoCo/Pt bilayer measured at various temperatures. The anomalous Hall resistivity $\rho_{xy}$ (H) at 116 K and below is negative and the loop is nearly square. The negative sign of $\rho_{xy}$ confirms that the $\mathbf{M}_{Co}$ is antiparallel to the applied field direction, thus leading to a negative value. The Ho sublattice, which is parallel to field, provides a reinforcing contribution to the total $\rho_{xy}$ due to its intrinsically negative AHE sign [44]. Here, both sublattices reverse simultaneously at the same coercive field resulting in a single sharp switching event with no intermediate steps. Further, no triple-loop structure is seen in the field range investigated here (≤5 tesla). As the temperature approaches $T_{comp}$ ≈ 135 K, the $\rho_{xy}$ (H) hysteresis broadens, becomes more tilted, and develops a distinct triple-loop structure. Above $T_{comp}$, the polarity of $\rho_{xy}$ (H) reverses (becoming positive), and the second loop in the first quadrant of the $\rho_{xy}$(H) curve shifts to higher fields. At T > 148 K, the loop centered at zero-field gradually recovers a square shape, and the minor loop is not seen up to a field of 5 tesla.

Likewise, the isothermal magnetization M(H) loops show marked changes in the 120 to 140 K range [Fig. 4(b)]. At 120 K, the M(H) exhibits a subtle multi-step reversal: the main switching is accompanied by a slight inflection (see inset of Fig. 4(b)), rather than a single sharp jump. Near

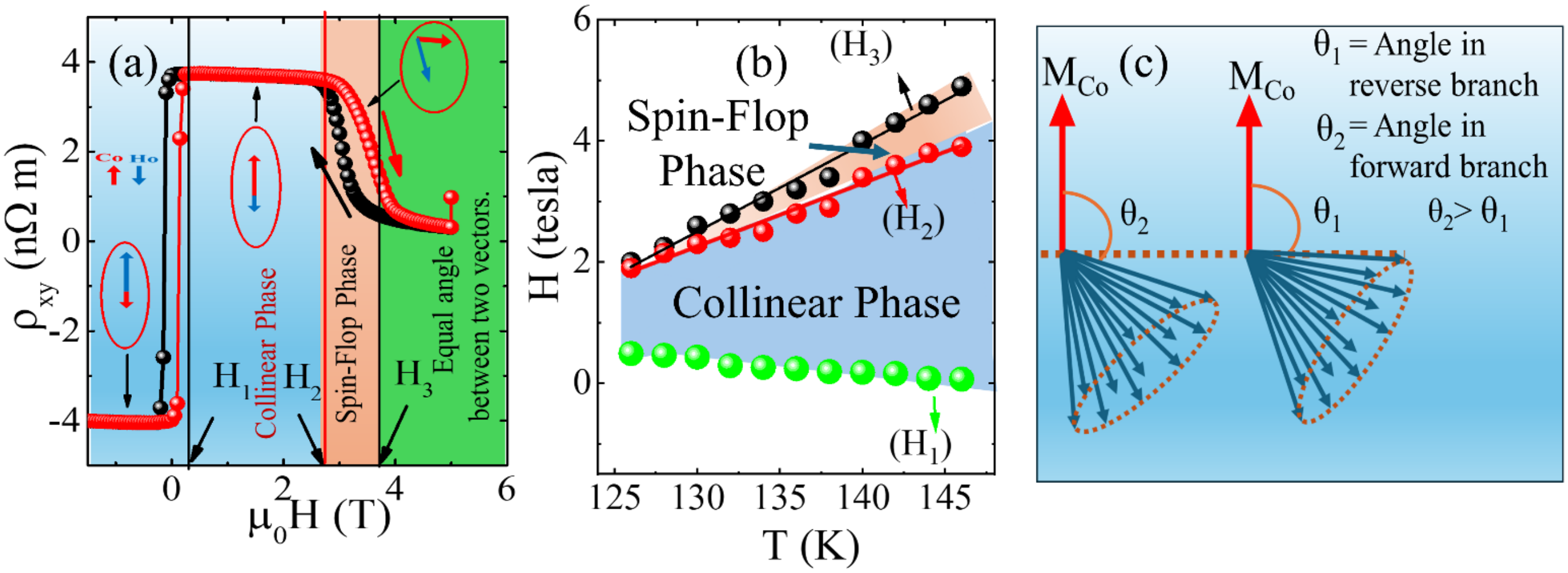

*Figure 5. (a) Hall resistivity shown for the positive field branches (0 to +5 T) at T = 140 K, highlighting the region between $H_1$, $H_2$ , and $H_3$ switching fields of the two-loop Hall signal and corresponding schematics of spin configurations. Red and blue arrows represent the relative orientation and magnitude of Co and Ho moments, respectively. (b) Representative Phase diagram of Pt/HoCo thin films constructed from the extraction of ($H_1$), ($H_2$ ), and $H_3$ switching fields of the triple-loop Hall signal. (c) Schematic representation of the fan-like splay of Ho magnetic moments forming a sperimagnetic state within the non-collinear phase. While the Co sublattice magnetization $M_{Co}$ remains aligned along the field direction, the Ho moments are distributed over*

*a broad angular cone, reflecting a noncollinear rare-earth spin configuration. The angles $\theta_1$and $\theta_2$correspond to the reverse and forward field-sweep branches, respectively, with $\theta_2 > \theta_1$, indicating hysteretic spin reorientation associated with field-driven transitions.*

the compensation point (≈135 K), the M(H) loop displays a pronounced double-step magnetization reversal, indicating two successive switching events associated with the two magnetic sublattices. Fig.4(c) compares the behaviors of $\rho_{xy}$(H) and M(H) at 140 K on an expanded scale, which both show a clear inflection around $\mu_0$(H) ≈2.6 tesla, which is indicative of a spin-flop-like transition in the system. For $\mu_0$(H) > 2.6 tesla, it is also noted that magnitudes of $\rho_{xy}$(H) and M(H) in the forward and reverse branches of the minor loop have opposite trends. A similar behavior was seen in the case of pure HoCo film (Fig. 3(c)) and explained on the basis of the relative orientations of $\mathbf{M}_{Co}$ and $\mathbf{M}_{Ho}$ as the field is cycled. Notably, for T < $T_{comp}$, the hysteresis loops are more complex: the multi-step features in M(H) and the negative $\rho_{xy}$ (H) signal indicate Ho-sublattice dominance in this regime. Similar to pure HoCo, these results can be understood within the framework of the two-sublattice model for anomalous Hall effect proposed by Okamoto and Miura for ferrimagnetic RE–TM alloys [16]. However, unlike the case of the HoCo film, the Pt/HoCo bilayer shows the persistence of the triple-loop structure over a broader temperature range. This suggests that the Pt interface modifies the inter-sublattice coupling or anisotropy, thereby extending the regime of competing sublattice dynamics [55, 56].

To further visualize and quantify the evolution of the triple-loop behavior across an extended temperature window, we constructed a H-T phase diagram from the anomalous Hall switching fields $H_1$ (corresponding to saturation of $\rho_{xy}$ beyond $H_c$), $H_2$ (where $\rho_{xy}$ starts decreasing), and $H_3$ where it again reaches a nearly constant value. As sketched in Fig. 5(a), for fields in the range of $H_1$ to $H_2$ $\mathbf{M}_{co}$ and $\mathbf{M}_{Ho}$ are in a collinear AF state. However, in the field range of $H_2$ and $H_3$ The angle Θ between these two vectors decreases from π but remains greater than π/2. However, its value for a given field in the forward and reverse branches of the loop are different.

If it is assumed that $\mathbf{M}_{Co}$ remains parallel to the external field for all H > $H_1$, $\mathbf{M}_{Ho}$ must rotate to account for the drop in $\rho_{xy}$ and increase in M(H). Further, the average angle Θ(H) in the reverse branch ($\Theta_1$(H)) must be smaller than the corresponding angle ($\Theta_2$(H)) in the forward branch to account for the higher moment seen in Fig. 4© and a lower $\rho_{xy}$ (Fig. 5(a)) on field reversal. The splay cone of Ho moments in Fig. 5(c) represents the sperimagnetic state of this rare-earth element [57]. Although this interpretation is consistent with the observed magnetization, anomalous Hall, angular-dependent transport behavior, and the constructed $H$-$T$ phase diagram, alternative scenarios involving complex field-dependent magnetic domain configurations cannot be completely excluded in the absence of direct microscopic probes of the sublattice-resolved magnetic structure. Within this framework, the field range between $H_2$ and $H_3$is associated with a spin-reorientation regime characterized by progressive canting of the Ho and Co sublattice moments away from the collinear ferrimagnetic configuration. The characteristic field $H_2$, extracted from the triple-loop Hall response, is used here as an estimate of the spin-flop field $H_{SF}$. Its temperature dependence, shown in the $H$-$T$ phase diagram [Fig. 5(b)], exhibits a non-monotonic behavior, with $H_{SF}$ reaching a minimum near the compensation temperature and increasing on either side away from compensation. Such behavior is consistent with spin-flop transitions in two-sublattice ferrimagnets, where the characteristic spin-flop field is governed by the competition between the effective inter-sublattice exchange interaction and magnetic anisotropy that stabilize the collinear ferrimagnetic state [11,24,51]. Near the compensation temperature, the reduced imbalance between the Ho and Co sublattice moments weakens the stability of the collinear ferrimagnetic configuration, leading to a lower $H_{SF}$, consistent with previous reports on RE–TM ferrimagnets [11,24]. Beyond the field H > $H_3$, the average angle Θ remains constant over the field range explored here. However, at a very large field, one would expect a complete FM coupling between $\mathbf{M}_{Co}$ and $\mathbf{M}_{Ho}$. Our dc magnetization measurement performed on the same set of samples supports this picture, as we see a lower moment in the AF state and a higher moment in the SF state. It is interesting to note that the presence of Pt underlayer makes the spin flop transition prominent compared to the behavior of the bare HoCo films. The Pt underlayer is susceptible to proximity magnetism and it may enhance interfacial anisotropy, both of which amplify the spin-flop window by stabilizing sublattice decoupling over a broader field range. Such behavior is consistent with earlier observations of multi-step Hall loops and side-wing features in Heavy metal/RE–TM heterostructure including Ta/TbFeCo[17] FeTb/[Co/Pt] [55] and Pt/$Tm_3Fe_5O_{12}$ [56], where interfacial anisotropy and sublattice competition similarly promote sequential reversal. This information, together with our results of dc magnetization measurements provide strong evidence that the observed switching in Pt/HoCo arises from an intrinsic, anisotropy-governed spin-flop mechanism characteristic of RE–TM ferrimagnets.

To further probe the spin reorientation behavior, we performed angle-dependent Hall measurements on the Pt/HoCo bilayer. Figure 6 b & c, respectively, show the behavior of $\rho_{xy}$ as a function of angle $\alpha$ and $\beta$ above and below the compensation temperature. The α and β rotations allow measurement of AHE when α or β are zero and planar Hall effect (PHE) for α,β = 90, which should be zero because of its SinϕCosϕ dependence for a collinear ferromagnet like permalloy [58]. Here ϕ is the azimuthal angle between the directions of current and magnetic field and ϕ = 0 corresponds to β = 0 and ϕ = 3π/2 to α = 0. The longitudinal component resistivity ($\rho_{xx}$) for α = 0 would yield spin Hall magnetoresistance (SMR) [58]. The angle-dependent measurements were performed at several temperatures ranging from 10 to 300 K, with the 130 K curve corresponding to the temperature closest to $T_{comp}$ (≈135 K). The curves are color-coded and vertically offset for clarity. Distinct changes in amplitude and angular symmetry of $\rho_{xy}$ are visible across the temperature range, reflecting the underlying evolution in magnetic anisotropy and sublattice alignment near compensation.

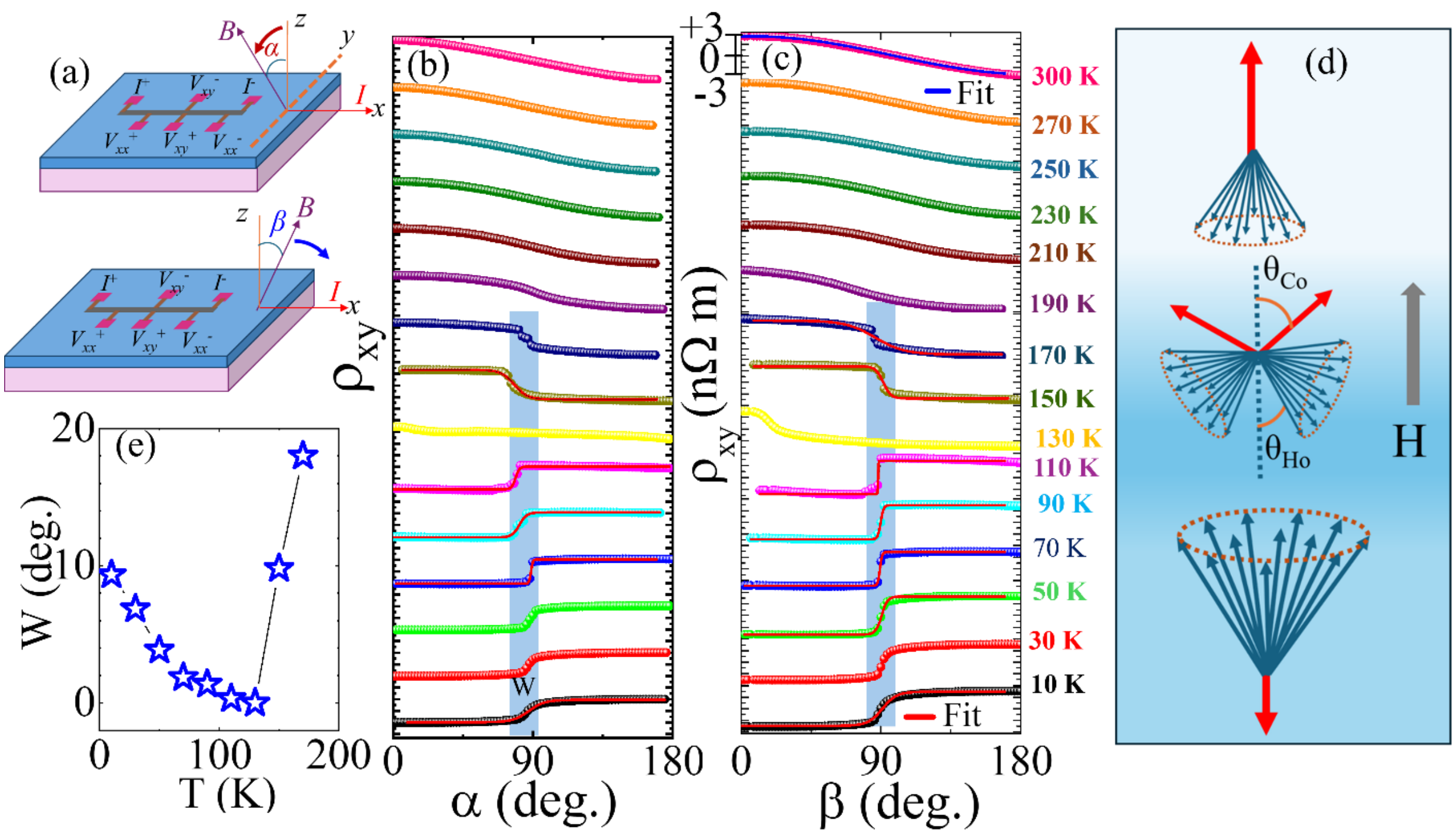

*Figure 6. (a) Schematic diagrams of the two rotation geometries: Top: field rotation in the yz-plane (α rotation); Bottom: rotation in the zx-plane (β rotation) (b) Angular dependence of the anomalous Hall resistivity ($\rho_{xy}$) (α rotation) at a fixed applied field of 2 tesla, for temperatures from 300 down to 10 K. (c) $\rho_{xy}$ (β rotation) under the same 2 tesla. All $\rho_{xy}$ data have been antisymmetrized to isolate the odd Hall component. Above $T_{comp}$, the Hall resistivity varies smoothly following a cosine dependence on the angle in both geometries, as expected for coherent magnetization rotation with the field. By contrast, at T below ~ 170 K the angular curves develop pronounced plateaus and abrupt jumps in ($\rho_{xy}$) signaling that the magnetization no more follows the field at all values of the angle. In this temperature regime, the $\rho_{xy}$ achieves full AHE value when the external field deviates slightly from the in-plane geometry (α and β = 90 degree), suggesting the preference of magnetization to stay out-of-plane. (d) Schematic illustration of the sublattice spin-canting mechanism underlying these effects: while the $M_{Co}$ (red arrow) flips by π at $T_{comp}$ , the $M_{Ho}$ (blue arrows) forms an umbrella pattern, with a wide open umrella of holmium moments at $T_{comp}$. This leads to a canted, non-collinear configuration of the two sublattices in the spin-flop state, which in turn produces the multi-step angular reorientation behavior observed in panels (b) and (c). The two cones in (d) schematically represent the sperimagnetic spin distributions of the Co and Ho sublattices, illustrating their non-collinear angular spread under an applied magnetic field. The different cone opening angles ($\theta_{Co}$ and $\theta_{Ho}$) reflect the unequal exchange stiffness and anisotropy of the transition-metal and rare-earth moments, leading to a net magnetization arising from vector summation rather than simple antiparallel alignment. (e) Temperature dependence of the anisotropy-dispersion parameter W extracted from sigmoidal fitting of the angular Hall data (β rotation). The corresponding fits are shown as red solid lines in Figs. 6(b and c). The width of this parameter is also highlighted by blue colors in the α– and β–dependent of $\rho_{xy}$ data shown in Figs. 6(b and c).*

Based on the standard anomalous Hall relation $\rho_{xy} = R_0 H_z + R_s M_z$, the angular Hall response is expected to follow a cosine dependence when the magnetization rotates coherently with the applied field [44]. The $\rho_{xy}$ of the HoCo/Pt bilayer shows a starkly different angular dependence above and below $T_{comp}$. For temperatures well above $T_{comp}$, where the TM sublattice dominates the magnetization, $\rho_{xy}$ varies smoothly with the angle of the external field following a cos (α) dependence. This reflects a spontaneous rotation of the net magnetization with the field. To quantitatively analyze the angular Hall response, the high-temperature data above $T_{\text{comp}}$ were fitted using the phenomenological form $\rho_{xy}(\alpha) = A + B\cos(\alpha - C)$, consistent with the anomalous Hall relation $\rho_{xy} \propto M_z$ for coherent magnetization rotation [44]. Here, $\alpha$ is the sample rotation angle with respect to the applied magnetic field direction, $A$ represents a constant background offset, $B$ is the amplitude of the anomalous Hall response associated with the out-of-plane magnetization component, and $C$ accounts for angular misalignment. In contrast, near and below $T_{\text{comp}}$, the angular Hall response develops a plateau–transition–plateau structure, indicating breakdown of coherent rotation due to strong anisotropy and sublattice competition. This regime is well described by a sigmoidal switching form, $\rho_{xy}(\alpha) = A\tanh[(\alpha - \alpha_c)/W]$, where $\alpha_c$ is the characteristic switching angle and $W$ represents the anisotropy-dispersion parameter associated with the angular broadening of the spin reorientation process, Such broadened switching behavior is qualitatively consistent with anisotropy-driven magnetization reversal and distributed switching expected in high-anisotropy magnetic systems [59, 60]. The extracted $W$ parameter exhibits a minimum near $T_{\text{comp}}$, indicating the sharpest and most collective spin reorientation near magnetic compensation, while larger $W$ values away from compensation suggest enhanced anisotropy dispersion and possible non-collinear spin canting. Here, the PHE is zero for $\alpha$ and $\beta = 0$, as expected from its $\phi$ dependence [58]. At temperatures below ≈ 170 K, the $\rho_{xy}$ does not change much as the angle ($\alpha$ or $\beta$) is increased towards 90 degrees. However, beyond $90^0$, the $\rho_{xy}$ changes polarity in a spring-like manner. This behavior indicates that at T < 170 K, the system develops a strong perpendicular anisotropy, and the magnetization vector refuses to follow the external field direction. But on crossing $90^o$, the magnetization flips by $180^0$, leading to a change in the sign of the $\rho_{xy}$.

A key factor underlying the contrasting behavior of the angular dependence of $\rho_{xy}$ above and below $T_{comp}$ is the magnetic anisotropy and coercivity of the rare-earth sublattice. Holmium has a large 4*f* orbital angular momentum (L = 6), resulting in strong spin–orbit coupling and a high anisotropy energy barrier. At $T < T_{comp}$, the HoCo exchange coupling is strong, which keeps the sublattices strictly antiparallel. Together, the large anisotropies of Ho pins the joint magnetization along the easy axis (out of the film plane). This appears to be the reason why the $\rho^{AHE}{}_{xy}$ (H) remains insensitive to the variations in $\alpha$ and $\beta$ as long as these angles are less than $\pi/2$ and changes sign in a spring-like manner only when $\alpha$ and $\beta$ become greater than 90 degrees.

The multistep behavior of $\rho_{xy}{}^{AHE}$ at $\alpha$ and $\beta$ = 90 degrees and $T \sim T_{comp}$ seen in Fig. 6(b and c) is suggestive of a non-collinear spin structure whose flipping on field reversal is not coherent. This situation has been sketched in Fig.6(d). Unlike most rare-earth–transition-metal (RE–TM) ferrimagnets [54], the HoCo/Pt bilayer exhibits a distinctly sharp spin reorientation near the compensation temperature.

<u>III.d. Enhanced SMR and Suppressed OMR in HoCo/Pt Heterostructures</u>

Figure 7 (a) and 7(b) respectively show the behavior of $\rho_{xx}$ of the bare HoCo film as a function of angles $\alpha$ and $\beta$ over the temperatures range of 10 to 300 K measured in a 2-tesla field. The changes in $\rho_{xx}(\alpha)$ with the angle $\alpha$ reflect the behavior of spin magnetoresistance (SMR), which arises due to spin current reflection at the interface [61-64], as well as the changes in resistance as the orbital motion of charge carriers and hence the scattering is affected as the field goes from out- of -plane to in-plane orientation. The $\rho_{xx}(\beta)$, on the other hand, reflects only the orbital part. The significant difference in the behaviors of $\rho_{xx}(\alpha)$ and $\rho_{xx}(\beta)$ seen in Fig. 7 suggests that a strong spin–orbit coupling in the bare HoCo film enables spin accumulation and backflow within its own conduction channels, producing an interfacial-like SMR response through the internal spin-mixing processes. To quantify the observed angular variations in magnetoresistance, we define the orbital magnetoresistance (OMR) as

$$\text{OMR} = [(\rho_{xx}(\beta = 90°) - \rho_{xx}(\beta = 0°)) / \rho_{xx}(\beta = 0°)] \times 100$$
$$= [(\rho_{\parallel} - \rho_{\perp}) / \rho_{\perp}] \times 100 \ldots\ldots\ldots\ldots\ldots\ldots\ldots (1)$$

where $\rho_{\parallel}$ and $\rho_{\perp}$ are the longitudinal resistivities when the external field is aligned parallel and perpendicular to the current direction, respectively.

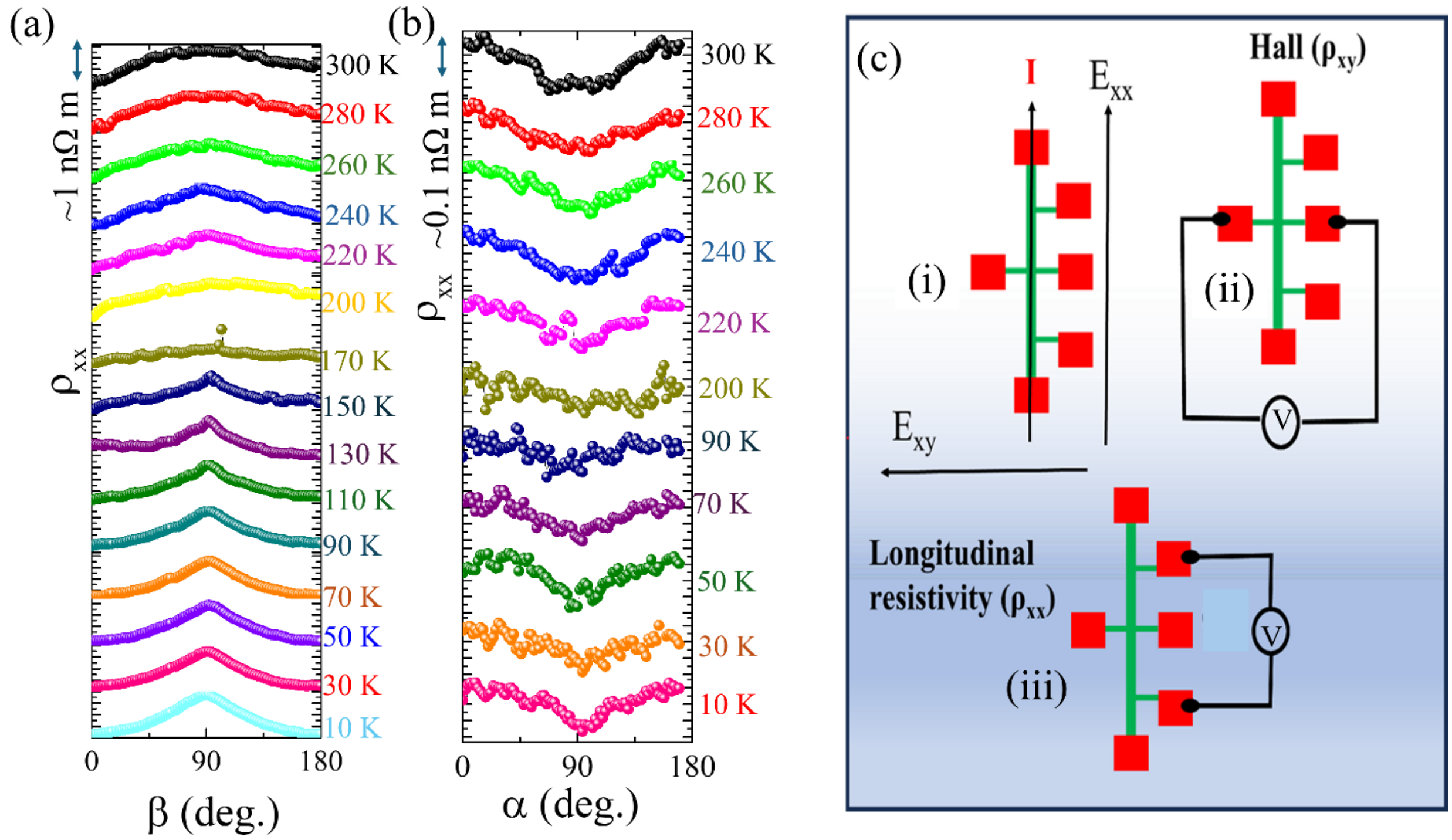

*Figure 7. Angular magnetoresistance of a $Ho_{22}Co_{78}$ (25 nm) film measured at 2 tesla. The (a) and (b) panels display the evolution of the OMR and SMR components as a function of angle β and α respectively over the temperature range 10 to 300 K. The rotation geometry for α and β is displayed in Figure 6 (a). (c) The schematic of the measurement configuration: (i) shows the orientations of the induced electric field $E_{xx}$ and $E_{xy}$, (ii) illustrates the transverse (Hall) voltage measurement used to obtain $\rho_{xy}$; and (iii) depicts the longitudinal voltage geometry used to measure $\rho_{xx}$.*

In the same manner the spin Hall magnetoresistance (SMR) is defined as:

$$SMR = [(\rho_{xx} (\alpha = 90°) - \rho_{xx} (\alpha = 0°)) / \rho_{xx} (\alpha = 0°)] \times 100$$
$$= (\Delta\rho / \rho_0) \times 100 \quad \text{......................} \quad (2)$$

Although the SMR is negative (Fig. 7b) we present the absolute value of the SMR, i.e., $|\Delta\rho|/\rho_0 \times 100$ in Fig. 8d.

The longitudinal electric fields associated with OMR and SMR, when current density $J_x$ is applied along the x-direction, are given by [63]:

$$E_{xx} (OMR) = [\rho\perp + (\rho\| - \rho\perp) \sin^2(\beta)] \times J_x \quad (3)$$

$$E_{xx} (SMR) = [\rho_0 + \Delta\rho \sin^2(\alpha)] \times J_x \quad (4)$$

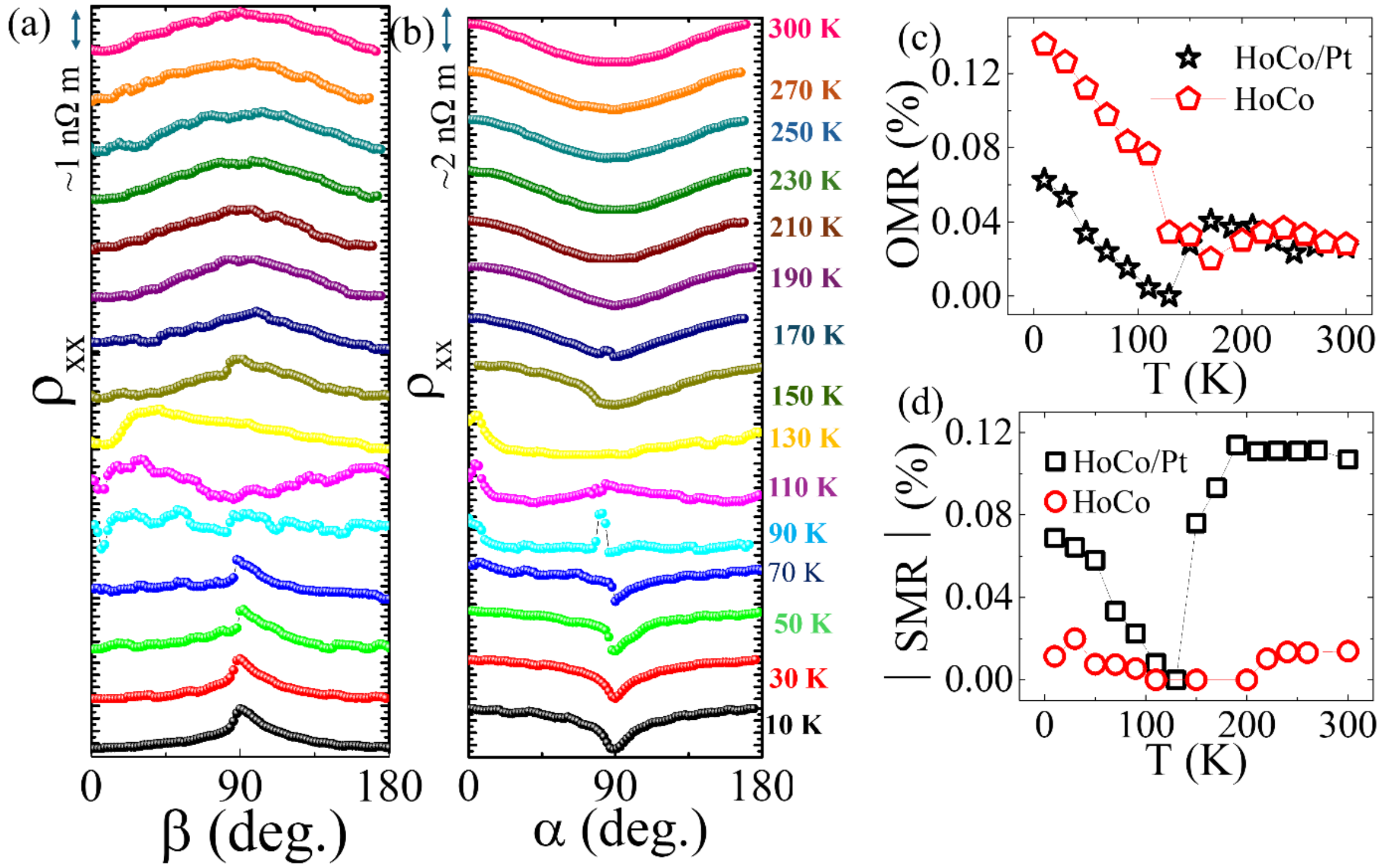

*Figure 8: Angular magnetoresistance of the HoCo/Pt film. The (a) and (b) panels display the evolution of SMR (α) and the OMR (β) components, respectively, as a function of rotating angle over the temperature range 10 to 300 K. The rotation geometry for α and β is displayed in Figure 6 (a). (c) Temperature dependence of the AMR ratio for the HoCo/Pt bilayer (Red Pentagon), compared to the pure HoCo film (black star). The bilayer's OMR amplitude is suppressed to nearly zero at the $T_{comp}$(≈130 K). The single-layer HoCo shows a larger OMR ratio below the compensation. (d) Temperature dependence of ratio in the Pt/Ho–Co bilayer (black rectangle) compared to the pure Ho–Co film (Red circle). The bilayer exhibits a significantly enhanced SMR compared to that of single layer Ho-Co.*

The Pt/HoCo bilayer exhibits a significantly enhanced SMR and a comparatively reduced OMR compared to the single HoCo film (Fig. 7). As shown in Fig 8(d), the SMR of the Pt/HoCo bilayer is ≈ 0.07% at 10 K followed by a drop to a nearly zero value at $T_{comp}$ ≈ 135 K and then a rise to ≈ 0.12% on reaching ~ 200 K. Beyond 200 K, the SMR is constant. By contrast, the single-layer HoCo shows only a small, nearly temperature-independent signal of ≈ 0.015% from ~ 0 – 200 K, followed by a subtle increase to ≈ 0.02% at 300 K. The significantly larger SMR in Pt/HoCo is due to spin Hall currents generated in the Pt layer and their interfacial exchange with HoCo [62-65]. Specifically, the SMR scales with spin accumulation at the Pt/HoCo interface and depends on the relative orientation of that spin polarization with respect to the local HoCo moments, as well as on the interfacial spin-mixing conductance and the spin polarization of HoCo's conduction electrons [62]. In metallic HoCo, the spin current generated in Pt is expected to be partially absorbed at the Pt/HoCo interface through spin-dependent reflection and interfacial spin-transfer processes [61-64]. Near $T_{\mathrm{comp}}$, the antiparallel Ho and Co sublattices do not simply eliminate the SMR through cancellation of the net magnetization; rather, their competing contributions can reduce the effective interfacial spin transparency and suppress the SMR amplitude despite the persistence of finite sublattice moments [63, 64]. Within the spin Hall magnetoresistance framework, the measured SMR ratio $\frac{\Delta\rho}{\rho_0}$ reflects the efficiency of spin-current absorption at the magnetic interface, which may be described in terms of an effective temperature-dependent spin-mixing conductance, $G_{\uparrow\downarrow}^{\mathrm{eff}}(T)$, rather than by the net magnetization alone [59–62]. The observed asymmetry of the SMR above and below $T_{\mathrm{comp}}$ suggests unequal interfacial coupling of spin currents to the Co and Ho sublattices, together with possible non-collinear spin configurations at low temperature [63,64]. Although a fully quantitative extraction of $G_{\uparrow\downarrow}^{\mathrm{eff}}(T)$ would require sublattice-resolved interfacial coupling parameters beyond the scope of the present work, the observed temperature dependence is consistent with a compensation-driven evolution of interfacial spin transparency in ferrimagnetic/heavy-metal bilayers [63,64], and future thickness- and composition-dependent studies should further clarify the interfacial origin of the observed SMR behavior. Further, the higher value of SMR at $T > T_{comp}$ is consistent with a crossover from Ho-dominated state below $T_{comp}$, where Co 3d carrier spins are antiparallel to the net moment, resulting in reduced spin mixing to Co-dominated state above $T_{comp}$ enabling enhanced spin-current reflection/absorption [64]. Complementary room-temperature Brillouin light scattering (BLS) measurements (Fig. S4 of the Supplementary Information) reveal well-defined, field-dependent spin-wave modes consistent with ferrimagnetic magnon dynamics, thereby supporting the presence of efficient interfacial spin angular momentum transfer relevant to spin pumping and spin-current transport in the Pt/HoCo bilayer.

The OMR of Pt/HoCo as shown in Fig. 8 (c) is only ≈ 0.07% at 10 K. It becomes nearly zero at $T_{comp}$ and then recovers to ≈ 0.03% at ≈ 200 K followed by a constant value thereafter. In pure HoCo, the OMR is larger at low temperature (≈ 0.13% at 10 K) and decays as T approaches $T_{comp}$, becoming comparable to the bilayer above compensation. The reduced OMR amplitude in the bilayer is consistent with current shunting into Pt and interfacial exchange that moderates the orbital-driven anisotropy originating from Ho $4f$ moments [66]. To further support this interpretation, temperature-dependent resistivity measurements on the HoCo film and Pt/HoCo bilayer are provided in the Supplementary Information (Fig. S6). The HoCo layer exhibits substantially higher resistivity than the Pt/HoCo bilayer, consistent with significant current shunting through the conductive Pt layer, which reduces the relative contribution of HoCo to the total transport response. However, the sharp suppression of OMR near $T_{\mathrm{comp}}$ cannot be explained by current shunting alone, since the resistivity evolution is comparatively smooth and does not exhibit a corresponding anomaly near compensation. This suggests that sublattice compensation and the associated reduction of magnetization-dependent orbital scattering play a dominant role in suppressing the OMR near $T_{\mathrm{comp}}$. Under our 2 T saturating field, the sublattices remain collinear across the measured temperature range, and no sign changes are observed [64]. Notably, the near-zero OMR at $T_{comp}$ indicates that the anisotropic magnetoresistance contributions from the Ho and Co sublattices cancel out when their magnetizations are equal and opposite. Moreover, the stark difference between the bilayer and single-layer OMR demonstrates that introducing a heavy-metal layer can substantially diminish a ferrimagnet's orbital magnetoresistance both by diverting a portion of the electric current into the nonmagnetic layer and thus reducing the magnetic scattering of charge carriers, and by exerting interfacial exchange that partially quenches

HoCo's intrinsic orbital anisotropy. Taken together, these results illustrate a general principle: in ferrimagnet/Pt bilayers, the heavy metal predominantly couples to the transition-metal sublattice of the ferrimagnet, so the spin Hall magnetoresistance remains appreciable as long as the transition-metal sublattice carries a moment, even if the oppositely polarized rare-earth sublattice cancels the net magnetization [65]. Interfacial effects such as spin-dependent scattering and the magnetic proximity effect may introduce secondary modifications to the magnetoresistance; for instance, Pt can acquire a slight induced magnetization that tracks the dominant Ho–Co sublattice, which can influence the resistivity [17,18, 49]. Overall, however, the trend is clear: adding a Pt layer robustly amplifies the spin-current-induced magnetoresistance (SMR) while concurrently attenuating the OMR of HoCo layer. This interplay yields the distinctive magnetotransport behavior of the HoCo/Pt heterostructure, especially around $T_{comp}$, and establishes this system as an ideal platform for studying spin transport in a nearly compensated magnetic system.

## IV. SUMMARY

This study demonstrates that HoCo ferrimagnetic alloys and their Pt-based heterostructures exhibit rich magnetotransport phenomena near the magnetic compensation point, most notably the emergence of triple hysteresis loops in the anomalous Hall effect linked to a field-induced spin-flop transition between antiferromagnetically coupled Co and Ho sublattices. Angle-dependent Hall measurements reveal spring-like behavior, where $\rho_{xy}$ remains pinned until a critical angle is exceeded, reflecting abrupt sublattice reorientation due to the high anisotropy of Ho. Interfacing of HoCo with Pt lowers its compensation temperature, enhances net magnetization through PIM, broadens the spin-flop regime, while also amplifying spin Hall magnetoresistance and suppressing orbital magnetoresistance near compensation. Importantly, the enhanced magnetization observed in the Pt/HoCo bilayer cannot be explained solely by conventional proximity-induced Pt moments, suggesting additional interfacial effects. Furthermore, the temperature dependence of the extracted spin-flop field reveals a minimum near the compensation temperature, consistent with reduced stability of the collinear ferrimagnetic state due to competing inter-sublattice exchange and magnetic anisotropy.

Overall, our work establishes that magnetic sublattice compensation in HoCo leads to novel magneto-transport signatures, including triple-loop Hall hysteresis, spin-flop metamagnetic transitions, and abrupt (spring-like) angular magnetization flips, which can be controlled and amplified by heavy-metal interfaces. Interfacial spin–orbit coupling and proximity magnetism (as introduced by Pt) provide additional knobs to tune the effective anisotropy and sublattice torque balance, thereby extending the regime of multi-step switching and enhancing spin current effects. These findings deepen our understanding of how competing 3d–4f sublattices behave under external fields and currents and demonstrate that Pt/HoCo heterostructures are a compelling platform for exploring spin transport in nearly compensated ferrimagnets. The ability to sustain sizable spin Hall signals and engineered triple-loop magnetization reversals at the compensation point could be exploited in future low-moment spintronic devices, offering high-speed magnetization dynamics with minimal stray fields and efficient spin-torque response.

## SUPPLEMENTARY INFORMATION

The Supplementary Information presents detailed RBS analysis verifying the multilayer composition (Fig. S1), temperature-dependent out-of-plane magnetization hysteresis loops identifying the compensation point (Figs. S2 and S3) through minimum in saturation magnetization, and Brillouin light scattering measurements revealing ferrimagnetic magnon dynamics (Fig. S4) that support the spin-pumping-driven transport results discussed in the main text. Figs. S5 and S6 show the Hall and magnetization of the Pt/$Co_{72}Ho_{28}$, and the temperature-dependent resistivity of Pt/$Co_{78}Ho_{22}$ and HoCo films, respectively.

## ACKNOWLEDGMENTS

This research is funded by the Air Force Office of Scientific Research under Grant No. FA9550-19-1-0082. Partial support has also come from the United States Department of Defense, Grant No. W911NF2120213. Structural and morphological characterizations, including XRD, XRR, AFM, and MFM were performed at the Air Force Research Laboratory, Dayton, Ohio, funded by Air Force Office of Scientific Research Grant No. LRIR 23RXCOR001. RBS measurements were conducted at the DEVCOM Army Research Laboratory, Aberdeen Proving Ground, Maryland.

## DATA AVAILABILITY

The data that support the findings of this study are available from the corresponding author upon reasonable request.

References

[1] S. K. Kim, G. S. D. Beach, K.-J. Lee, T. Ono, T. Rasing, and H. Yang, "Ferrimagnetic spintronics", Nat. Mater. **21**, 24 (2022).

[2] Y. Zhang et al., "Ferrimagnets for spintronic devices: From materials to applications", Appl. Phys. Rev. **10**, 11301 (2023).

[3] J. Finley and L. Liu, "Spintronics with compensated ferrimagnets", Appl. Phys. Lett. **116**, 110501 (2020).

[4] T. Jungwirth, X. Marti, P. Wadley, and J. Wunderlich, "Antiferromagnetic spintronics", Nat. Nanotechnol. **11**, 231 (2016).

[5] A. V Kimel and M. Li, "Writing magnetic memory with ultrashort light pulses", Nat. Rev. Mater. **4**, 189 (2019).

[6] T. Satoh, S.-J. Cho, R. Iida, T. Shimura, K. Kuroda, H. Ueda, Y. Ueda, B. A. Ivanov, F. Nori, and M. Fiebig, "Spin Oscillations in Antiferromagnetic NiO Triggered by Circularly Polarized Light", Phys. Rev. Lett. **105**, 77402 (2010).

[7] F. Schlickeiser, U. Ritzmann, D. Hinzke, and U. Nowak, "Role of Entropy in Domain Wall Motion in Thermal Gradients", Phys. Rev. Lett. **113**, 97201 (2014).

[8] S. Woo et al., "Current-driven dynamics and inhibition of the skyrmion Hall effect of ferrimagnetic skyrmions in GdFeCo films", Nat. Commun. **9**, 959 (2018).

[9] J. Becker, A. Tsukamoto, A. Kirilyuk, J. C. Maan, T. Rasing, P. C. M. Christianen, and A. V Kimel, "Ultrafast Magnetism of a Ferrimagnet across the Spin-Flop Transition in High Magnetic Fields", Phys. Rev. Lett. **118**, 117203 (2017).

[10] M. B. Jungfleisch, W. Zhang, and A. Hoffmann, "Perspectives of antiferromagnetic spintronics", Phys. Lett. A **382**, 865 (2018).

[11] M. D. Davydova, K. A. Zvezdin, J. Becker, A. V Kimel, and A. K. Zvezdin, "T–H phase diagram of rare-earth–transition-metal alloys in the vicinity of the compensation point", Phys. Rev. B **100**, 64409 (2019).

[12] J.-L. Liu, P.-B. He, and M.-Q. Cai, "Current-driven spiral domain wall in a ferrimagnet near the magnetization compensation point", Phys. Rev. Res. **4**, 23253 (2022).

[13] G. Sala and P. Gambardella, "Ferrimagnetic Dynamics Induced by Spin-Orbit Torques", Adv. Mater. Interfaces **9**, 2201622 (2022).

[14] H. Ratajczak and I. Gošciańska, "Hall hysteresis loops in the vicinity of compensation temperature in amorphous HoCo films", Phys. Status Solidi **62**, 163 (1980).

[15] T. Chen and R. Malmhäll, "Anomalous hysteresis loops in single and double layer sputtered TbFe films", J. Magn. Magn. Mater. **35**, 269 (1983).

[16] K. Okamoto and N. Miura, "Hall effect in a RE-TM perpendicular magnetic anisotropy film under pulsed high magnetic fields," Phys. B Condens. Matter **155**, 259 (1989).

[17] M. D. Davydova et al., "Unusual field dependence of the anomalous Hall effect in Ta/Tb-Fe-Co", Phys. Rev. Appl. **13**, 34053 (2020).

[18] A. Chanda, J. E. Shoup, N. Schulz, D. A. Arena, and H. Srikanth, "Tunable competing magnetic anisotropies and spin reconfigurations in ferrimagnetic Fe 100− x Gd x alloy films", Phys. Rev. B **104**, 94404 (2021).

[19] K. Chen, D. Lott, F. Radu, F. Choueikani, E. Otero, and P. Ohresser, "Observation of an atomic exchange bias effect in DyCo4 film", Sci. Rep. **5**, 18377 (2015).

[20] C. Luo, H. Ryll, C. H. Back, and F. Radu, "X-ray magnetic linear dichroism as a probe for non-collinear magnetic state in ferrimagnetic single layer exchange bias systems", Sci. Rep. **9**, 18169 (2019).

[21] M. Amatsu, S. Honda, and T. Kusuda, "Anomalous hysteresis loops and domain observation in Gd-Fe Co-evaporated films", IEEE Trans. Magn. **13**, 1612 (1977).

[22] C. Xu, Z. Chen, D. Chen, S. Zhou, and T. Lai, "Origin of anomalous hysteresis loops induced by femtosecond laser pulses in GdFeCo amorphous films", Appl. Phys. Lett. **96**, 92514 (2010).

[23] S. Esho, "Anomalous Magneto-optical Hysteresis Loops of Sputtered Gd-Co Films", Jpn. J. Appl. Phys. **15**, 93 (1976).

[24] A. Pogrebna, K. Prabhakara, M. Davydova, J. Becker, A. Tsukamoto, T. Rasing, A. Kirilyuk, A. K. Zvezdin, P. C. M. Christianen, and A. V Kimel, "High-field anomalies of equilibrium and ultrafast magnetism in rare-earth--transition-metal ferrimagnets", Phys. Rev. B **100**, 174427 (2019).

[25] H. Wu et al., "Spin-Orbit Torque Switching of a Nearly Compensated Ferrimagnet by Topological Surface States", Adv. Mater. **31**, 1901681 (2019).

[26] D. Chen, Y. Xu, S. Tong, W. Zheng, Y. Sun, J. Lu, N. Lei, D. Wei, and J. Zhao, "Noncollinear spin state and unusual magnetoresistance in ferrimagnet Co-Gd", Phys. Rev. Mater. **6**, 14402 (2022).

[27] T. A. Peterson, A. P. McFadden, C. J. Palmström, and P. A. Crowell, "Influence of the magnetic proximity effect on spin-orbit torque efficiencies in ferromagnet/platinum bilayers", Phys. Rev. B **97**, 20403 (2018).

[28] C. Swindells, H. Głowiński, Y. Choi, D. Haskel, P. P. Michałowski, T. Hase, P. Kuświk, and D. Atkinson, "Proximity-induced magnetism and the enhancement of damping in ferromagnetic/heavy metal systems", Appl. Phys. Lett. **119**, 152401 (2021).

[29] H. L. Wang, C. H. Du, Y. Pu, R. Adur, P. C. Hammel, and F. Y. Yang, "Scaling of Spin Hall Angle in 3d, 4d, and 5d Metals from Y3Fe5O12/Metal Spin Pumping", Phys. Rev. Lett. **112**, 197201 (2014).

[30] J. Bass and W. P. Pratt, "Spin-diffusion lengths in metals and alloys, and spin-flipping at metal/metal interfaces: an experimentalist's critical review", J. Phys. Condens. Matter **19**, 183201 (2007).

[31] C. Swindells, B. Nicholson, O. Inyang, Y. Choi, T. Hase, and D. Atkinson, "Proximity-induced magnetism in Pt layered with rare-earth--transition-metal ferrimagnetic alloys", Phys. Rev. Res. **2**, 33280 (2020).

[32] S. Catalano, J. M. Gomez-Perez, M. X. Aguilar-Pujol, A. Chuvilin, M. Gobbi, L. E. Hueso, and F. Casanova, "Spin Hall Magnetoresistance Effect from a Disordered Interface", ACS Appl. Mater. Interfaces **14**, 8598 (2022).

[33] M. Althammer, A. V. Singh, T. Wimmer, Z. Galazka, H. Huebl, M. Opel, R. Gross, and A. Gupta, "Role of interface quality for the spin Hall magnetoresistance in nickel ferrite thin films with bulk-like magnetic properties", Appl. Phys. Lett. **115**, 92403 (2019).

[34] F. Hellman et al., Interface-induced phenomena in magnetism, Rev. Mod. Phys. **89**, 25006 (2017).

[35] F. Wilhelm, P. Poulopoulos, H. Wende, A. Scherz, K. Baberschke, M. Angelakeris, N. K. Flevaris, and A. Rogalev, "Systematics of the Induced Magnetic Moments in 5d Layers and the Violation of the Third Hund's Rule", Phys. Rev. Lett. **87**, 207202 (2001).

[36] J. Vogel, A. Fontaine, V. Cros, F. Petroff, J.-P. Kappler, G. Krill, A. Rogalev, and J. Goulon, "Structure and magnetism of Pd in Pd/Fe multilayers studied by x-ray magnetic circular dichroism at the Pd L2,3 edges", Phys. Rev. B **55**, 3663 (1997).

[37] D. M. Burn, T. P. A. Hase, and D. Atkinson, "Focused-ion-beam induced interfacial intermixing of magnetic bilayers for nanoscale control of magnetic properties", J. Phys. Condens. Matter **26**, 236002 (2014).

[38] D. Weller, J. Stöhr, R. Nakajima, A. Carl, M. G. Samant, C. Chappert, R. Mégy, P. Beauvillain, P. Veillet, and G. A. Held, "Microscopic Origin of Magnetic Anisotropy in Au/Co/Au Probed with X-Ray Magnetic Circular Dichroism", Phys. Rev. Lett. **75**, 3752 (1995).

[39] M. Suzuki, N. Kawamura, H. Miyagawa, J. S. Garitaonandia, Y. Yamamoto, and H. Hori, "Measurement of a Pauli and Orbital Paramagnetic State in Bulk Gold Using X-Ray Magnetic Circular Dichroism Spectroscopy", Phys. Rev. Lett. **108**, 47201 (2012).

[40] Knížek, K., & Jirák, Z., "Magnetic proximity effects at a ferrite–5 d-metal interface: GGA+ U calculations", Phys. Rev. B **108**, 35145 (2023).

[41] N. Reynolds, P. Jadaun, J. T. Heron, C. L. Jermain, J. Gibbons, R. Collette, R. A. Buhrman, D. G. Schlom, and D. C. Ralph, "Spin Hall torques generated by rare-earth thin films", Phys. Rev. B **95**, 64412 (2017).

[42] T. Seifert et al., "Terahertz Spin Currents and Inverse Spin Hall Effect in Thin-Film Heterostructures Containing Complex Magnetic Compounds", SPIN **07**, 1740010 (2017).

[43] F. Radu, R. Abrudan, I. Radu, D. Schmitz, and H. Zabel," Perpendicular exchange bias in ferrimagnetic spin valves", Nat. Commun. **3**, 715 (2012).

[44] R. Nepal, V. Sharma, J. T. Sadowski, and R. C. Budhani, "Anomalous and planar Hall effects in $Co_{1-x}Ho_x$ thin films across the magnetic sublattice compensation temperature", J. Appl. Phys. **138**, 43902 (2025).

[45] G. Garreau, E. Beaurepaire, M. Farle, and J.-P. Kappler, "Second- and fourth-order anisotropy constants near the spin reorientation transition in Co/Ho thin films," Europhys. Lett. **39**, 557 (1997).

[46] S. S. Nandra and P. J. Grundy, "Lorentz microscopy of the magnetic domain structure in amorphous rare earth-cobalt films evaporated at normal incidence", Phys. Status Solidi **41**, 65 (1977).

[47] W. Amamou et al., "Magnetic proximity effect in $Pt/CoFe_2O_4$ bilayers", Phys. Rev. Mater. **2**, 11401 (2018).

[48] F. Wilhelm et al., "Layer-resolved magnetic moments in Ni/Pt multilayers", Phys. Rev. Lett. **85**, 413 (2000).

[49] N. Nakajima, T. Koide, T. Shidara, H. Miyauchi, H. Fukutani, A. Fujimori, K. Iio, T. Katayama, M. Nývlt, and Y. Suzuki, "Perpendicular magnetic anisotropy caused by interfacial hybridization via enhanced orbital moment in Co/Pt multilayers: Magnetic circular x-ray dichroism study", Phys. Rev. Lett. **81**, 5229 (1998).

[50] A. Mukhopadhyay, S. Koyiloth Vayalil, D. Graulich, I. Ahamed, S. Francoual, A. Kashyap, T. Kuschel, and P. S. Anil Kumar, "Asymmetric modification of the magnetic proximity effect in Pt/Co/Pt trilayers by the insertion of a Ta buffer layer", Phys. Rev. B **102**, 144435 (2020).

[51] A. V Svalov, A. S. Rusalina, E. V Kudyukov, V. N. Lepalovskij, E. A. Stepanova, A. A. Yushkov, V. O. Vas'kovskiy, and G. V Kurlyandskaya, "Temperature and field dependencies of the magnetization of ferrimagnetic Gd-Co films: Chemical inhomogeneity or spin-flop transition", J. Non. Cryst. Solids **640**, 123116 (2024).

[52] Y. Wang, C. Li, H. Zhou, J. Wang, G. Chai, and C. Jiang, "Unusual anomalous Hall effect in the ferrimagnetic GdFeCo alloy", Appl. Phys. Lett. **118**, 71902 (2021).

[53] A. Johnson, E. Negusse, V. Sharma, D. Anyumba, D. McAlmont, and R. C. Budhani, "Sublattice magnetization driven anomalous Hall resistance of FeCoGd amorphous films", AIP Adv. **10**, 115214 (2020).

[54] T. Fu, S. Li, X. Feng, Y. Cui, J. Yao, B. Wang, J. Cao, Z. Shi, D. Xue, and X. Fan, "Complex anomalous Hall effect of CoGd alloy near the magnetization compensation temperature", Phys. Rev. B **103**, 64432 (2021).

[55] C. Schubert, B. Hebler, H. Schletter, A. Liebig, M. Daniel, R. Abrudan, F. Radu, and M. Albrecht, "Interfacial exchange coupling in Fe-Tb/[Co/Pt] heterostructures", Phys. Rev. B **87**, 54415 (2013).

[56] Y. K. Liu, H. F. Wong, S. M. Ng, C. L. Mak, and C. W. Leung, "Interfacial $Tm^{3+}$-moment–driven anomalous Hall effect in $Pt/Tm_3Fe_5O_{12}$ heterostructures", J. Magn. Magn. Mater. **501**, 166454 (2020).

[57] J. M. D. Coey, J. Chappert, J. P. Rebouillat, and T. S. Wang, Magnetic Structure of an Amorphous Rare-Earth Transition-Metal Alloy, Phys. Rev. Lett. 36, 1061 (1976).

[58] T. McGuire and R. Potter, "Anisotropic magnetoresistance in ferromagnetic 3d alloys", IEEE Trans. Magn. **11**, 1018 (1975).

[59] Stoner, E. C., and E. P. Wohlfarth, Philos. Trans. R. Soc. London, Ser. A 240, 599 (1948).

[60] Hubert, A., and R. Schäfer, *Magnetic Domains: The Analysis of Magnetic Microstructures* Springer Science & Business Media, (1998).

[61] J. Kim, P. Sheng, S. Takahashi, S. Mitani, and M. Hayashi, "Spin Hall Magnetoresistance in Metallic Bilayers", Phys. Rev. Lett. **116**, 97201 (2016).

[62] N. Vlietstra, J. Shan, V. Castel, B. J. van Wees, and J. Ben Youssef, "Spin-Hall magnetoresistance in platinum on yttrium iron garnet: Dependence on platinum thickness and in-plane/out-of-plane magnetization", Phys. Rev. B **87**, 184421 (2013).

[63] W. Zhou, T. Seki, T. Kubota, G. E. W. Bauer, and K. Takanashi, "Spin-Hall and anisotropic magnetoresistance in ferrimagnetic Co-Gd/Pt layers", Phys. Rev. Mater. **2**, 94404 (2018).

[64] Y. Xu, D. Chen, S. Tong, H. Chen, X. Qiu, D. Wei, and J. Zhao, "Spin Polarization Compensation in Ferrimagnetic Co1-xTb1-xTbx/Pt Bilayers Revealed by Spin Hall Magnetoresistance", Phys. Rev. Appl. **14**, 34064 (2020).

[65] J. J. Bauer, P. Quarterman, A. J. Grutter, B. Khurana, S. Kundu, K. A. Mkhoyan, J. A. Borchers, and C. A. Ross, "Magnetic proximity effect in magnetic-insulator/heavy-metal heterostructures across the compensation temperature", Phys. Rev. B **104**, 94403 (2021).

[66] A. B. Shick, D. S. Shapiro, J. Kolorenc, and A. I. Lichtenstein, "Magnetic character of holmium atom adsorbed on platinum surface", Sci. Rep. **7**, 2751 (2017).

SUPPLIMENTARY INFORMATION

# Magneto-Transport and Spin-Reorientation in Pt/$Co_{78}Ho_{22}$ Heterostructures Near the Sublattice Compensation Temperature

Rajeev Nepal[1], Jose Flores[2], Aurain Seaton[1], Michael Newburger[2], John Derek Demaree[3] and Ramesh C Budhani[1*]

This Supplementary Information provides additional structural, magnetic, and dynamic characterization supporting the main article on spin-pumping-driven magneto-transport in Pt/$Co_{1-x}Ho_x$ heterostructures near the ferrimagnetic sublattice compensation temperature. Rutherford backscattering spectroscopy (Fig. S1) confirms the film composition and interfacial structure. Out-of-plane magnetization hysteresis loops (Figs. S2 and S3) reveal a minimum in saturation magnetization, identifying the compensation temperature. Brillouin light scattering (BLS) measurements (Fig. S4) show field-dependent spin-wave modes, confirming ferrimagnetic magnon dynamics relevant to spin pumping and spin transport. The BLS data was collected in a 180° backscattering geometry using a focused laser spot (~5 μm). The laser wavelength was 532 nm and the signal was collected and analyzed via 6 pass tandem Fabry-Perot interferometer. In all the measurements, the polarization of the analyzed light was 90° rotated with respect to the excitation polarization to select for the magnon signal. In the field range measured (~40-210 mT applied in the plane of the sample) we observe a clear peak which is resolvable on both the stokes and anti-stokes side, confirming its origin as the sample and not a spurious mode. Further, the mode frequency increases monotonically with applied field, consistent with a Kittel-like behavior of the acoustic ferrimagnetic magnon mode, confirming the magnetic nature of the signal. This represents an optical analogue to ferromagnetic resonance measurements, which confirms the presence of ferrimagnetic dynamics in the expected frequency regime for this composite system. We note that the peak width is comparable to other metallic ferrimagnetic alloys, indicating good quality material with reasonable uniformity and magnetic damping. Future studies could focus on analyzing the dispersion of this mode via angle dependent measurements, or analysis of the exchange stiffness and Dzyaloshinskii-Moriya interaction which are both extractable through additional BLS measurements but are outside the scope of this work.

In addition to the primary Pt/$Co_{78}Ho_{22}$ heterostructures discussed in the main text, supplementary magnetization and anomalous Hall measurements on a second Pt/$Co_{72}Ho_{28}$ sample with a compensation temperature near room temperature ($T_{\mathrm{comp}} \approx 285$ K) are presented in Fig. S5. These measurements show a polarity reversal of the anomalous Hall resistivity together with the emergence of a distinct triple-loop hysteresis structure near $T_{\mathrm{comp}}$, indicating complex magnetization reversal processes arising from competing ferrimagnetic sublattice interactions close to magnetic compensation. Figure S6 presents the temperature-dependent resistivity of the same Pt/$Co_{78}Ho_{22}$ heterostructure discussed in the main manuscript, together with the corresponding HoCo reference film. The comparison highlights the semiconducting-like transport behavior of the HoCo layer and the metallic transport response of the Pt/HoCo bilayer due to the conductive Pt overlayer and interfacial charge transport contributions.

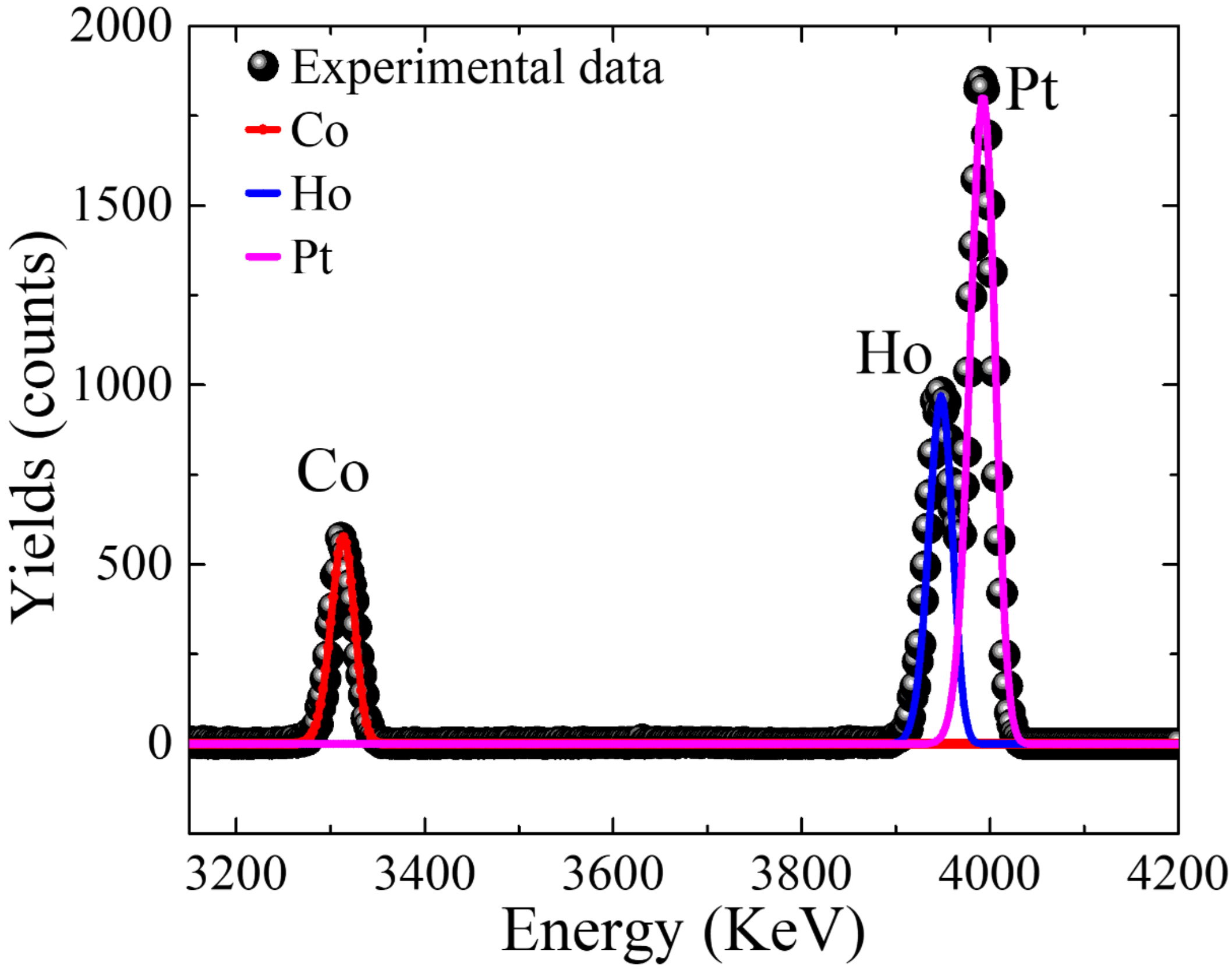

*Fig.S1. Rutherford backscattering spectroscopy (RBS) spectrum of the Pt/HoCo thin film measured with 4.365 MeV He ions at 175°. The experimental data are well reproduced by the simulation (Lines), showing distinct energy edges for Co, Ho, and Pt. The sharp high-energy peaks correspond to surface scattering from heavy Pt and Ho atoms, while the broader features reflect the underlying Co layer. RBS confirms the composition and interfacial structure of the multilayer. The areal densities of Ho and Co are determined to be* $4.1 \times 10^{16}$ *and* $1.49 \times 10^{17}$ *atoms/cm*$^2$*, respectively, corresponding to a Ho:Co ratio of approximately 1:3.6 (~21.6% Ho and ~78.4% Co).*

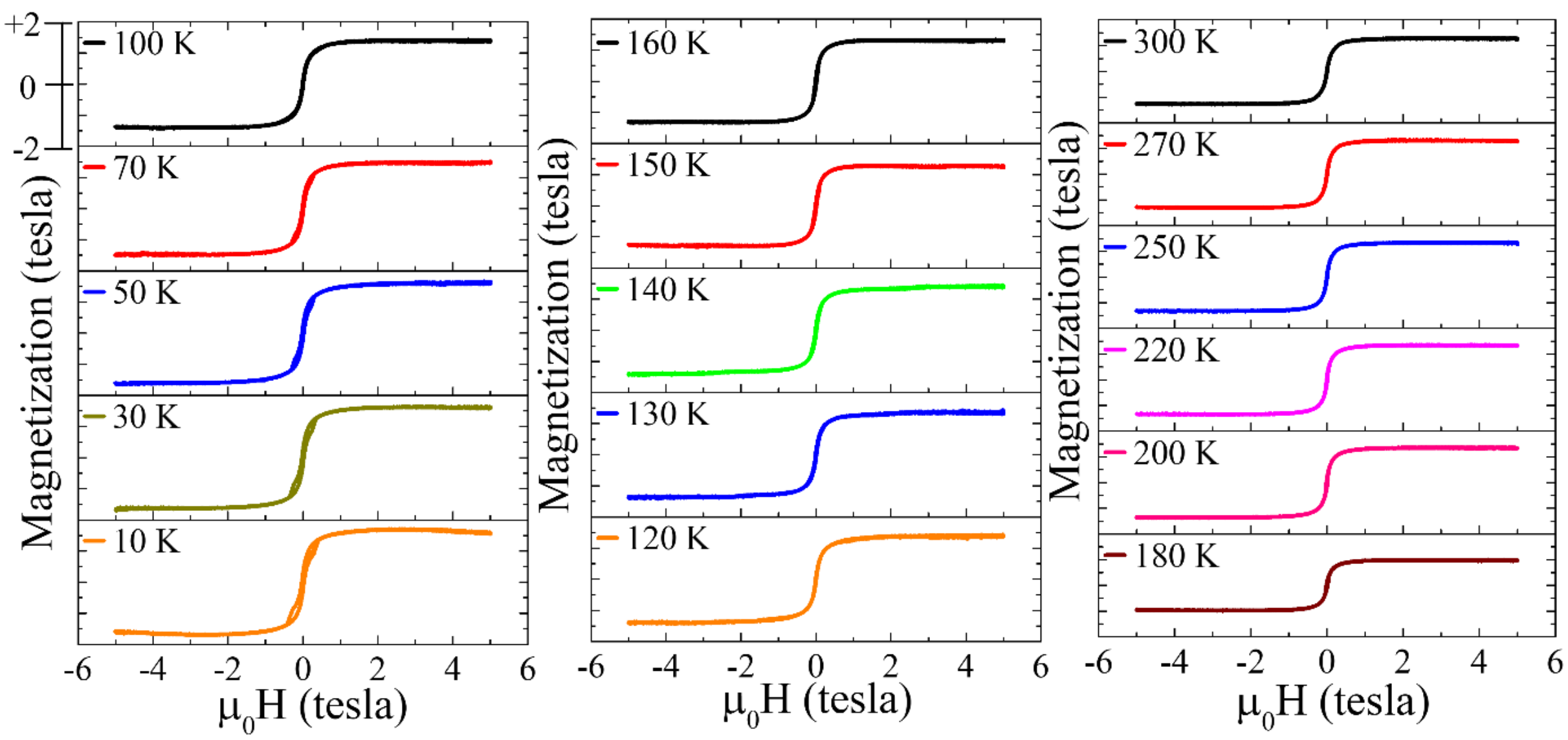

*Fig.S2. Out-of-plane magnetization hysteresis loops of Pt/HoCo thin films measured at various temperatures.*

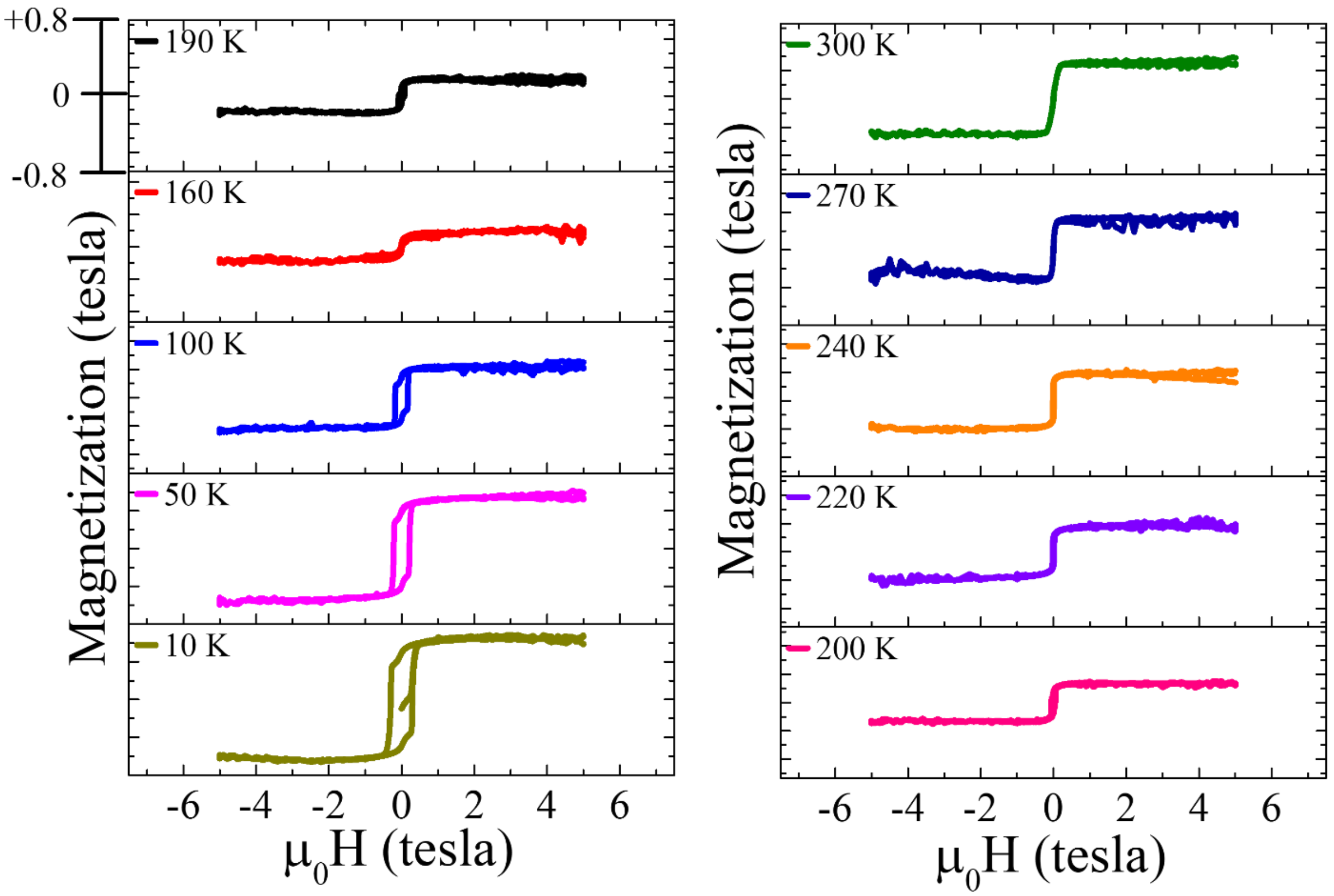

*Fig. S3. Out-of-plane magnetization hysteresis loops of the HoCo thin film measured at temperatures between 10 and 300 K. The saturation magnetization reaches a minimum at approximately at 190 K, indicating the magnetic compensation temperature of the film.*

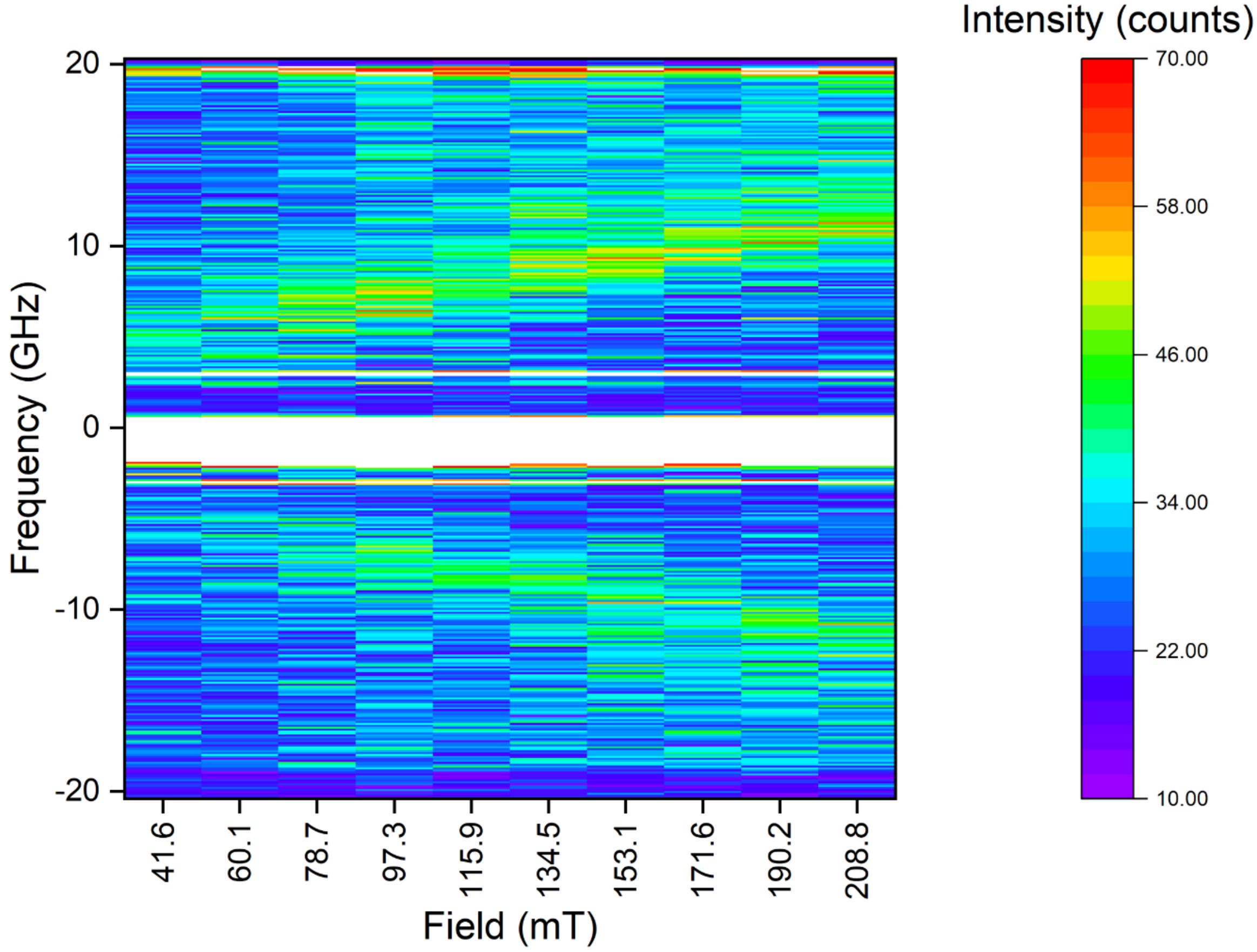

*Fig. S4. Brillouin light scattering (BLS) intensity map of the Pt–HoCo bilayer measured at room temperature. The color scale represents the BLS intensity (counts), plotted as a function of the applied in-plane magnetic field H (mT) and frequency (GHz).*

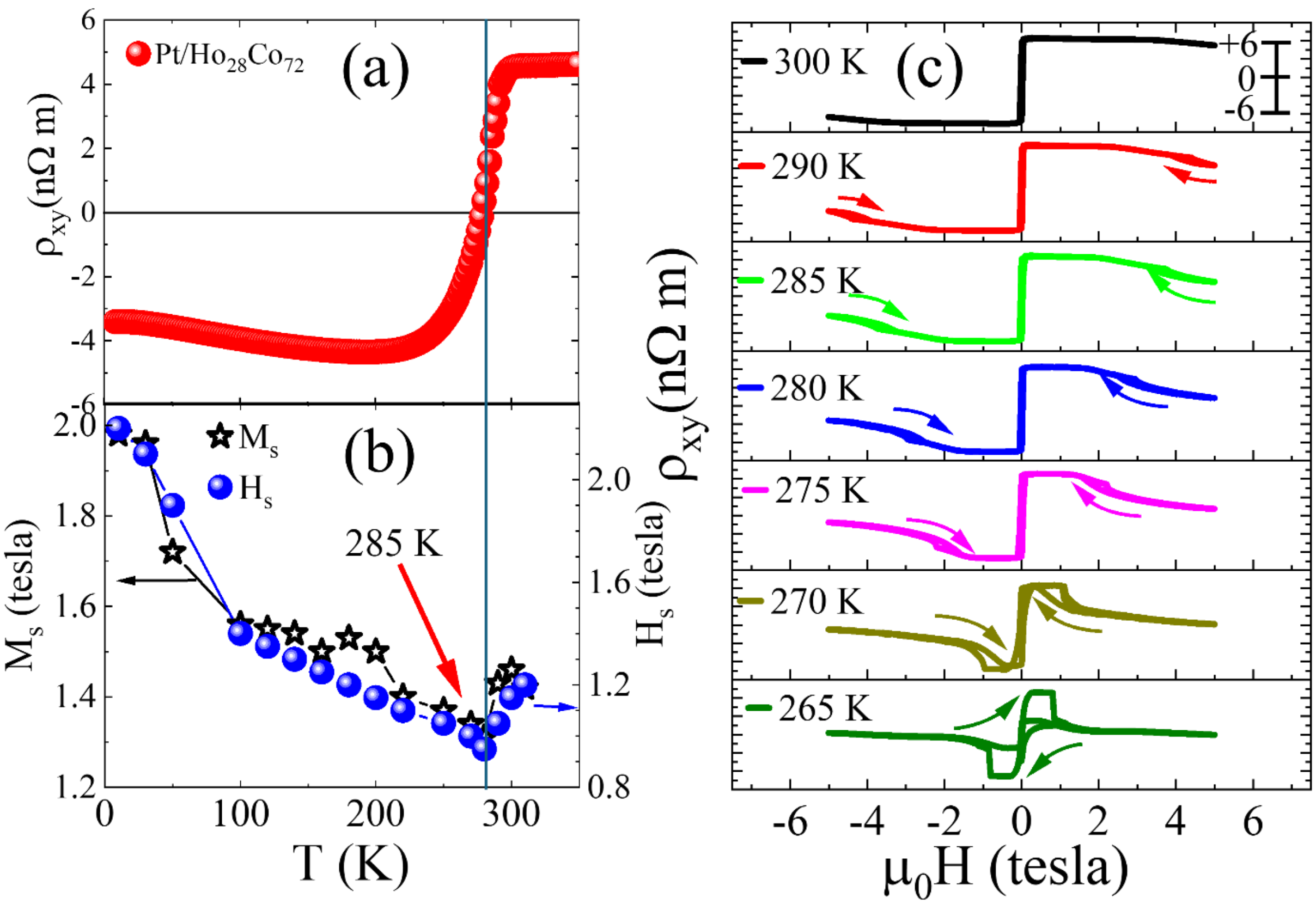

*Fig. S5. Magnetization and Hall resistivity measurements of the Pt/$Ho_{28}Co_{72}$ bilayer with the magnetic field applied perpendicular to the film plane (out-of-plane). (a) Hall resistivity $\rho_{xy}$ as a function of temperature measured at 2 T, showing a polarity reversal near the magnetic compensation temperature ($T_{comp} \approx 285$ K). $\rho_{xy}$ data were antisymmetrized to isolate the odd Hall component. (b) Temperature dependence of the saturation magnetization ($M_s$) and saturation field ($H_s$), where the minimum in $M_s$ identifies the compensation temperature. (c) Temperature-dependent anomalous Hall hysteresis loops $\rho_{xy}(H)$ measured between 265 and 300 K, showing the emergence of a distinct triple-loop structure near $T_{comp} \approx 285$ K.*

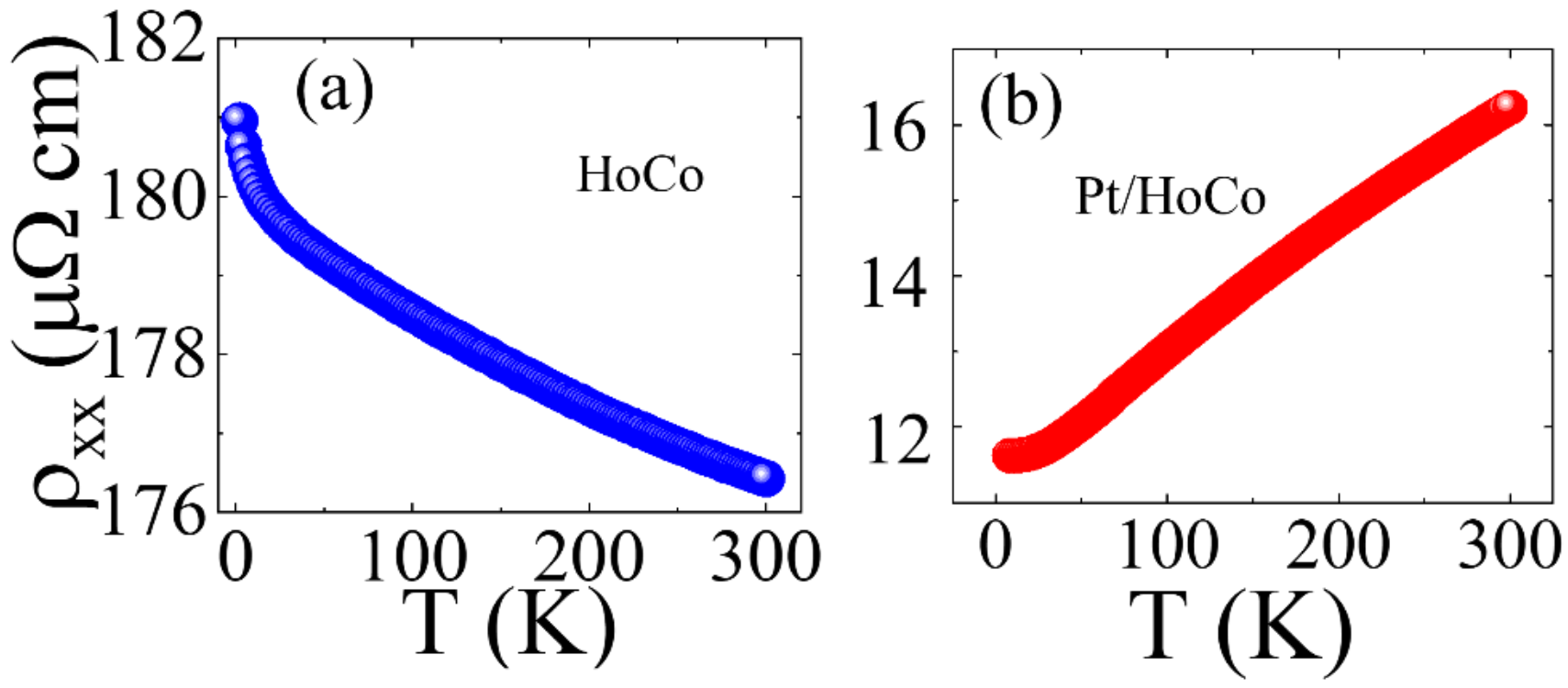

*Fig. S6. Temperature-dependent resistivity of the HoCo thin film and Pt/HoCo bilayer. The HoCo film exhibits semiconducting-like transport behavior, while the Pt/HoCo bilayer is distinctly metallic with reduced resistivity due to the conductive Pt layer and interfacial transport effects.*